\documentclass[11pt,a4paper]{article}

\usepackage{iftex}

\ifPDFTeX
    \usepackage[T1]{fontenc}
    \usepackage[utf8]{inputenc}
    \usepackage{lmodern}

    \input{glyphtounicode}
\else
    \usepackage{fontspec}
    \defaultfontfeatures{Ligatures=TeX}
\fi

\usepackage[USenglish]{babel}
\usepackage{microtype}

\usepackage[margin=1.4cm]{geometry}
\usepackage{setspace}
\usepackage{fancyhdr}
\usepackage{float}

\usepackage{amsmath}
\usepackage{amssymb}
\usepackage{amsfonts}
\usepackage{amsthm}
\usepackage{bm}
\usepackage{mathrsfs}
\usepackage{relsize}
\usepackage{slashed}
\usepackage{polynom}

\usepackage{graphicx}
\usepackage{pict2e}
\usepackage{overpic}
\usepackage{caption}
\usepackage{subcaption}
\usepackage{tikz}

\usepackage[table,xcdraw]{xcolor}
\usepackage{multirow}

\usepackage{stfloats}

\usepackage{enumitem}
\usepackage{listings}

\definecolor{dkgreen}{rgb}{0,0.6,0}
\definecolor{gray}{rgb}{0.5,0.5,0.5}
\definecolor{mauve}{rgb}{0.58,0,0.82}

\usepackage[round]{natbib}

\usepackage[
    colorlinks=true,
    linkcolor=blue,
    citecolor=black,
    urlcolor=black
]{hyperref}

\theoremstyle{plain}

\newtheorem{theorem}{Theorem}

\newtheorem{claim}[theorem]{Claim}

\theoremstyle{definition}

\theoremstyle{remark}

\newtheorem*{remarkunnumbered}{Remark}

\newcommand{\trp}{t_r^+}

\newcommand{\keywords}[1]{%
    \par\medskip
    \noindent\textbf{Keywords:} #1
    \par
}

\usepackage{authblk}
\begin{document}

\title{Mathematical Modeling and Optimization of Athletic Performance:
Tapering and Periodization}

\author[1]{David Ceddia\thanks{Corresponding author:
\href{mailto:d.ceddia@unimelb.edu.au}{d.ceddia@unimelb.edu.au}}}
\author[1]{Howard Bondell}
\author[1]{Peter Taylor}

\affil[1]{%
School of Mathematics and Statistics,
The University of Melbourne,
Melbourne, Australia
}

\date{}
\maketitle

\abstract{
We conduct a mathematical optimization of the training impulse profile to maximize performance for two seminal athletic performance models: the \citet{banister1975systems} Fitness--Fatigue Impulse Response Model and the \citet{busso2003variable} Variable Dose--Response Model. We discuss discrepancies in both the quantitative and qualitative aspects of the optimized training impulse profiles compared to what is physically plausible and relative to common training practices recommended in the empirical sports science literature, such as periodization and tapering. We analytically prove that, under the Banister formulation, the optimal training strategy always consists of sustained maximal training, followed by the complete cessation of training prior to the prioritized performance date, and therefore does not yield periodization, or tapering beyond this trivial form. This highlights a gap between theoretical prediction and empirically supported practice, underscoring the need for models that incorporate additional physiological effects. We then propose a set of illustrative and interpretable nonlinear modifications in the magnitude and time response to training in the Fitness--Fatigue Impulse Response model such that the optimized training impulse profile demonstrates these qualities.
}


\section{Introduction}

The first widely acknowledged quantitative model to map a training impulse profile to a change in athletic performance is the \citet{banister1975systems} Fitness--Fatigue Impulse Response model. Since 1975, this, and variants, have been applied to numerous sports and study participants, including: 
swimming \citep{banister1975systems,mujika1996modeled,hellard2006assessing,thomas2008model}, 
running \citep{morton1990modeling}, 
weight lifting \citep{busso1990systems}, 
cycling \citep{busso1991adequacy,busso1997modeling}, 
the hammer throw \citep{busso1994fatigue},
triathlons \citep{banister1999training,millet2002modeling},
gymnastics \citep{sanchez2013modeling} and 
judo \citep{agostinho2015perceived}.
In addition to using a Fitness--Fatigue Impulse Response model to predict the change in performance that would result from a particular training impulse profile, a natural question to ask would be what training impulse profile maximizes the improvement in performance by a prioritized date. Indeed, it is worth investigating whether the optimal training impulse profiles derived from these models are capable of reflecting the qualitative behavior described in contemporary training theory or that which is typically seen in elite athlete training impulse profiles documented in the observational sports science literature -- see, for example, the case study of three elite marathon runners presented in \citet{stellingwerff2012case}. To our knowledge, the qualitative structure of the optimal training impulse profiles implied by these models has not yet been systematically characterized. That is, in this study, we investigate whether optimized Fitness--Fatigue Impulse Response models recommend training impulse profiles that qualitatively display periodization and tapering.

Periodization can be defined as a training impulse profile that exhibits purposeful, non-random variation at characteristic temporal scales, for example, daily, weekly, or monthly. For a general overview of periodization, including traditional and block variations, see \citet{issurin2010new}; for a complementary, contemporary review with illustrations of the concept of periodization, see \citet{turner2011science}. 
For the purposes of this study, any training impulse profile that shows clear cyclical trends such that regular peaks and troughs can be identified in the training impulse profile at characteristic frequencies is said to display periodization. We recognize that this is a simplified definition of ``periodization.'' The mathematical construction of training impulses does not, however, account for factors such as training type or intent, and therefore cannot incorporate these aspects that are typically included in the conventional meaning of the term.

Another formal training impulse profile pattern often employed to maximize performance on a prioritized date is tapering \citep{vachon2021effects}. This is the practice of successively reducing the training impulse in the period leading up to the prioritized performance date in an attempt to recover from the fatigue generated during the training preceding the taper. The contemporary meta-review by \citet{wang2023effects}, which builds upon \citep{bosquet2007effects}, concludes ``current evidence suggests that a $\leq$21-day taper, in which training volume is progressively reduced by 41–60\% without changing training intensity or frequency, is an effective tapering strategy.''

In this paper, we optimize the training impulse profile for two seminal quantitative athletic performance models, the \citet{banister1975systems} Fitness--Fatigue Impulse Response model and the \citet{busso2003variable} Variable Dose--Response model. We then address questions of whether the optimal training impulse profiles are plausibly realistic, and whether they qualitatively display behaviors that are in line with empirically derived guidelines for periodization and tapering. For cases of discrepancies, there can be two possible reasons:
\begin{enumerate}
    \item the models do not incorporate all the relevant effects,
    \item the commonly understood training behavior is not optimal.
\end{enumerate}
Assuming that the discrepancies are due to the first point, we present illustrative and interpretable modifications to the Fitness--Fatigue Impulse Response model that produce periodization and tapering in the training impulse profile that maximizes performance. Note, the question of which modifications are required to produce a given realistic training impulse profile is a context-dependent problem and is beyond the resolution of this study.


In Section \ref{Sec:Background}, we introduce the \citet{banister1975systems} Fitness--Fatigue Impulse Response model and the \citet{busso2003variable} Variable Dose--Response model, which employ the concept of a training impulse.
In Section \ref{Sec:ExistingModeloptimization}, we establish model-derived training impulse constraints and numerically optimize the Banister and Busso models, explore their prediction for the optimum training impulse profile which maximizes performance and discuss shortcomings. 
In Section \ref{Sec:Model Modifications}, we systematically introduce interpretable nonlinear modifications to the Fitness--Fatigue model to elicit trends in the optimized training impulse profile that are consistent with common elite athlete training practices, such as tapering and periodization. 
In Section \ref{sec:Discussion}, we discuss broader implications of our work and offer several suggestions for future research. 
In Section \ref{sec:Conclusion}, we present our conclusions.

\section{Existing Athletic Performance Models and Training Impulse Framework} \label{Sec:Background}

\subsection*{Banister et al.'s Fitness--Fatigue Impulse Response Model} \label{Sec:fitness-fatigue models}

The Fitness--Fatigue model proposed by \citet{banister1975systems,calvert1976systems} and subsequently refined in \citet{morton1990modeling,busso1994fatigue} expresses the performance $P$ in terms of $w$, a continuous, non-negative, one dimensional training variable, which is a function of continuous time $t$, via the equation
\begin{align} \label{Eq:OGBanisterContinuous}
    P(t) = P_0 + \int_{0}^{t} w(t')g(t-t')dt',
\end{align}
where
\begin{align*}
g(t) = k_1 \exp{\left( - \frac{t}{\tau_1} \right)} - k_2 \exp{\left( - \frac{t}{\tau_2} \right)}
\end{align*}
is known as the transfer function, $P_0$ is a baseline performance, $k_1$ is the magnitude of fitness gains from a unit of training, $k_2$ is the magnitude of fatigue incurred by a unit of training, and $\tau_1$ and $\tau_2$ are time constants that control the decay of the fitness and fatigue response, respectively. Throughout, we consider the regime in which performance improvements are positive, which implies $0<k_1<k_2$ and $0<\tau_2 <\tau_1$. If performance improvements were negative, the corresponding parameter relations would become $k_2<k_1<0$ and $0<\tau_2 <\tau_1$.

\begin{figure}[ht!]
     \centering
     \begin{subfigure}{0.32\textwidth}
         \centering
         \includegraphics[width=\textwidth]{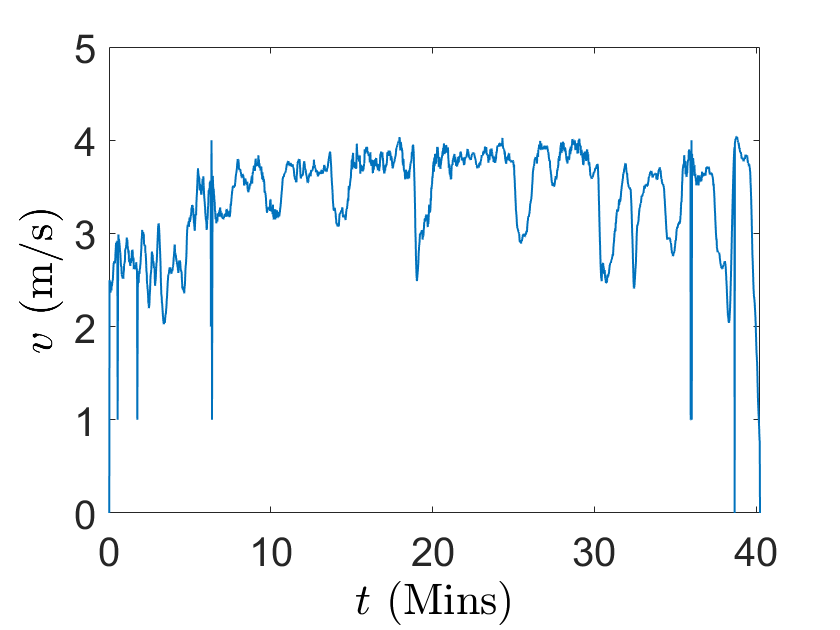}
         \caption{ }
         \label{subfig:g01}
     \end{subfigure}
     \begin{subfigure}{0.32\textwidth}
         \centering
         \includegraphics[width=\textwidth]{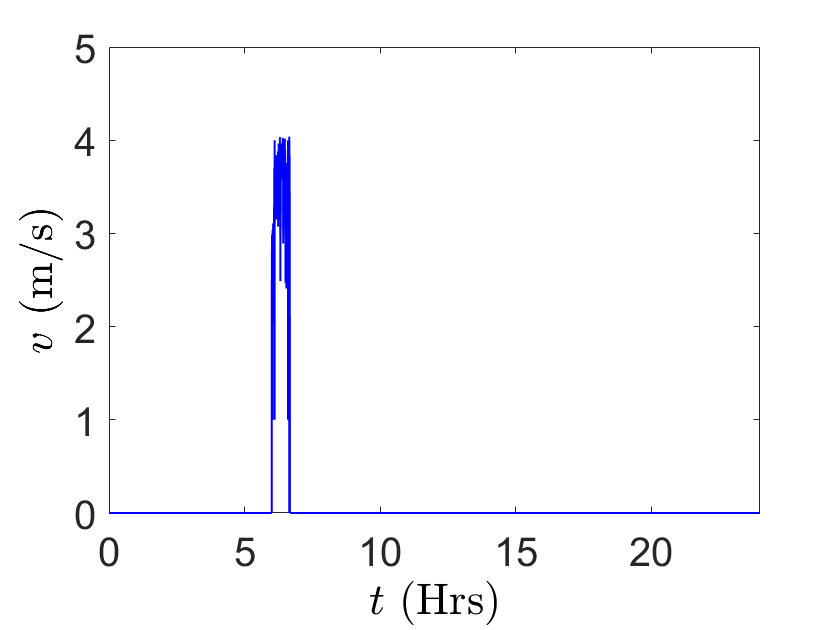}
         \caption{ }
         \label{subfig:g02}
     \end{subfigure}
     \begin{subfigure}{0.32\textwidth}
         \centering
         \includegraphics[width=\textwidth]{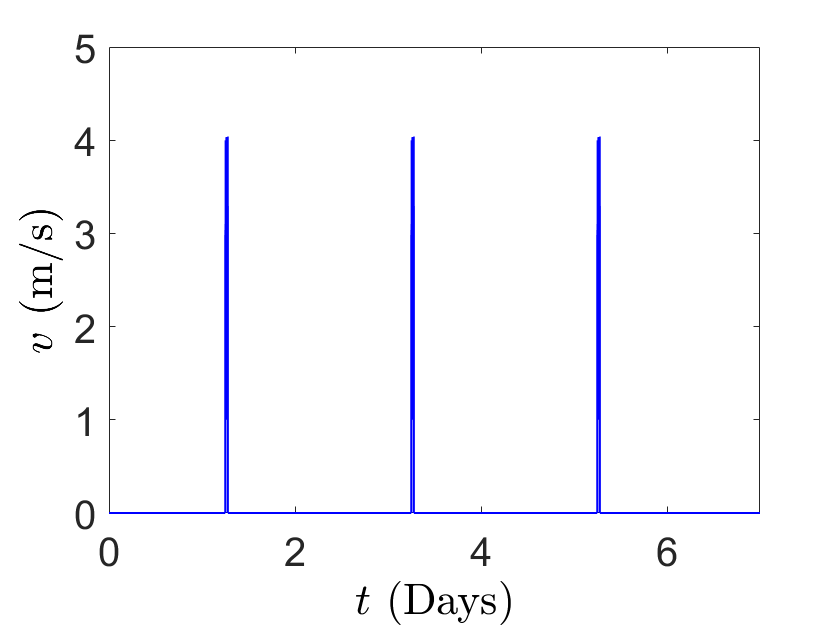}
         \caption{ }
         \label{subfig:g03}
     \end{subfigure}
        \caption{Illustration of the training impulse approximation. (a) Training velocity as a function of time. (b) Same, but viewed on the timescale of a day. (c) Same, but repeated on Tuesday, Thursday and Saturday, viewed on the timescale of a week.}
        \label{fig:ImpulseApprox}
\end{figure}

An important simplifying step is to notice that the training time frame is relatively small compared to the resting time frame, which allows a continuous time training session to be summarized by a discrete, one-dimensional training impulse $w$ that quantifies its impact. For an illustrative comparison of the training to resting time frames, see Fig.~\ref{fig:ImpulseApprox}. In this, we see a single training session viewed on a timescale of its duration, the same training session viewed on the timescale of a day, and the same training session, repeated Tuesday, Thursday and Saturday, viewed on the timescale of a week. As such, the integrand in \eqref{Eq:OGBanisterContinuous} is mostly zero and the non-zero contributions can be captured as a discrete sum of one-dimensional training impulses. 

In the original work of \citet{banister1975systems}, the authors studied the change in performance of a swimmer and quantified the training impulse using a 3-tier weighting system based on intensity for each 100m swum. Warm-up intensity was awarded 1 arbitrary training unit (ATU) per 100m swum, low intensity was awarded 2 ATUs per 100m swum and high intensity was awarded 3 ATUs per 100m swum. A training impulse for each session was then calculated by summing up the weights for each 100m swum. Additionally, weight-training was taken into account via the estimate that 500 lightweight pulls was considered equivalent to 1000m of high-intensity swimming or 30 ATUs. An assortment of subsequent definitions have since come along, some such being: the heart rate (HR) system $w = D (\Delta \text{HR Ratio}) \exp(b\, \Delta\text{HR Ratio}) $, where $D$ is the training duration and $(\Delta \text{HR Ratio}) = (\text{HR}_{\text{av}} -\text{HR}_{\text{rest}})/(\text{HR}_{\text{max}}-\text{HR}_{\text{rest}})$ and $b = 1.92$ for men and $b = 1.67$ for women \citep{morton1990modeling}; or to apply weightings of 1,2,3,4,5 in a five-HR-zone, linear system \citep{edwards1994heart}; or weights of 1,2,3,5,8 in a five-lactate-zone, approximately exponential system \citep{mujika1996modeled}; or weights of 1,2,3 in a three-HR-zone, linear system (1 $<$aerobic threshold, 2 between thresholds, 3 $>$anaerobic threshold) \citep{lucia1999heart}. 
The particular training impulse formula chosen will be context-dependent, and a discussion of which is most appropriate is outside of the scope of this paper. We simply note that training is often summarized by a single number. Our interest lies in exploring the mathematical ramifications of this. Note, with respect to terminology, we will refer to a single peak as a `training impulse', and use `training impulse profile' to refer to a sequence or collection of training impulses.

The discrete version of \eqref{Eq:OGBanisterContinuous} is
\begin{align} \label{Eq:OGBanisterDiscrete}
    P(n) = P_0 + \sum_{i=1}^{n-1} w(i) \left[  k_1 \exp{\left( - \frac{(n-i)}{\tau_1} \right)}- k_2 \exp{\left( - \frac{(n-i)}{\tau_2} \right)} \right]
\end{align}
where $P(n)$ is the predicted performance on day $n \in \mathbb{Z}^{+}$, $w(i)$ is the training impulse for a given day and $i \in \{1,\dots,n-1\}$ sums over all prior contributions to the day of interest. Physiologically speaking, $\tau_1$ and $\tau_2$ control the time frames associated with detraining and recovery, respectively. 
Henceforth, we will refer to \eqref{Eq:OGBanisterDiscrete} as the Banister model, the cumulative positive impulse response, $F(n) = \sum_{i=1}^{n-1} k_1 w(i) \exp{( -(n-i)/\tau_1)}$, as the fitness and the cumulative negative impulse response, $f(n) = \sum_{i=1}^{n-1} k_2 w(i) \exp{( -(n-i)/\tau_2)}$, as the fatigue. Additionally, we emphasize that $w(i)$ is a training impulse and the exponential decay is the impulse response.

It is worth noting that not all physiological phenomena take place on time scales appropriate to be accurately captured by a daily training impulse, for example, a warm-up. Moreover, with the definition of training impulse, there now exists an ambiguity in the discernment of intensity and duration, which complicates the reverse process of going from an optimized training impulse profile, to an actual implementable training program. A potential resolution is to design training sessions that match the prescribed training impulse while adhering to training intensity distribution guidelines \citep{sun2025recent}. In any case, the construction of a daily training impulse does capture phenomena that persist from one training session to the next, and it can aid in the planning of large-scale training structures, such as periodization and tapering strategies.

Given the model in \eqref{Eq:OGBanisterDiscrete}, one can calculate two timepoints of interest. The first is the time
\begin{align} \label{eq:Time-to-return}
    t_{r} = \left( \frac{\tau_1 \tau_2 }{\tau_1 -\tau_2} \right) \ln \left( \frac{k_2}{k_1} \right)
\end{align}
until the impulse response reaches a break-even point, after which a positive return is made. This is found by taking the positive and negative impulse responses, and solving for the time at which the two are equal. From this, it is practical to define
\begin{align} \label{eq:trp}
    \trp = \lfloor t_r \rfloor + 1
\end{align}
where $\lfloor t_r \rfloor$ is the floor of $t_r$ in \eqref{eq:Time-to-return}, as this is the number of days required for the impulse response to first make a positive return. 
Given that $0<k_1<k_2$, $t_r$ will be some positive time in the future. The second time of interest,
\begin{align} \label{eq:Time-to-peak}
    t_{p} = \left( \frac{\tau_1 \tau_2 }{\tau_1 -\tau_2} \right) \ln \left( \frac{k_2 \tau_1}{k_1 \tau_2} \right)
\end{align}
is the time it takes the impulse response to reach a peak. This is found by taking the time derivative of the combined fitness--fatigue impulse response and solving for the time $t_p$ where it is equal to zero. 

Applying the Banister model to time-series training impulse and performance data involves estimating the model parameters. Typically, this is done by fitting the model to minimize the Residual Sum of Squares (RSS) between predicted and actual performance measurements for a given set of training impulse data \citep{busso1991adequacy,busso1997modeling}. To give an idea as to how this model performs in an applied context, consider the findings of \citet{hellard2006assessing}, which applied the Banister model to nine swimmers and found that the change in performance was related to the training impulse distribution for all participants. A mean value of 0.79 for the coefficient of determination, $r^2$, was obtained for the nine swimmers, with a standard deviation of 0.13 and a range of [0.61, 0.97].

\subsection*{Busso's Variable Dose--Response Model} \label{Sec:Variable Dose Response}

A significant development on the Banister model is the \citet{busso2003variable} Variable Dose--Response model, which introduces a nonlinear fatigue response. This model has a similar structure to the Banister model in equation \eqref{Eq:OGBanisterDiscrete}, but captures the idea that the fatigue magnitude, $k_2$, varies as a function of $i$ and depends on the recent training. The model takes the form
\begin{align} \label{Eq:VDR}
    P(n) &= P_0 + \sum_{i=1}^{n-1} w(i)\left[ k_1 \exp{\left( - \frac{(n-i)}{\tau_1} \right)}- k_2(i) \exp{\left( - \frac{(n-i)}{\tau_2} \right)} \right],
\end{align}
where 
\begin{align*}
    k_2(i) &= \sum_{j=1}^{i-1} w(j) k_3 \exp{\left( - \frac{(i-j)}{\tau_3} \right)},
\end{align*}
denotes the variable magnitude of fatigue incurred by a unit of training on day $i$; $k_3$ and $\tau_3$ are the magnitude and decay time constant, respectively, that control the variable fatigue-magnitude response. In other words, the fatigue generated at the current time due to a training impulse is governed by a memory of recent training impulses, the amount and duration of which are determined by $k_3$ and $\tau_3$.
In a contemporary study comparing several statistical tools applied to training data and performance outcomes, this model was used as the benchmark athletic performance model from which to gauge the various modeling approaches' effectiveness \citep{imbach2022training}. Henceforth, we will refer to \eqref{Eq:VDR} as the Busso model.

\section{Optimization of the Banister and Busso Models} \label{Sec:ExistingModeloptimization}

\subsection*{The Banister Model}

For the purposes of optimization, we underscore the importance and practicality of the simplifying assumption introduced by \citet{banister1975systems}, that both training impulse and performance are expressed as a function of discrete days where training is summarized by a one-dimensional variable $w$.

Consider the training impulse profile leading up to day $n$, a prioritized performance time point in the future, denoted by $\bm{w} = (w(1), \ldots ,w(n-1))$. Using the Banister model defined in \eqref{Eq:OGBanisterDiscrete}, we can obtain a prediction for an athlete's performance on day $n$. A natural question for sports scientists to ask is, ``What training impulse profile should an athlete adopt to maximize their performance on day $n$?'' Mathematically speaking, this translates to solving the optimization problem
\begin{align} \label{eq:BanisterOptimizationProblem}
    \max_{\bm{w}} ~ P_0 + \sum_{i=1}^{n-1} w(i) \left[  k_1 \exp{\left( - \frac{(n-i)}{\tau_1} \right)}- k_2 \exp{\left( - \frac{(n-i)}{\tau_2} \right)} \right].
\end{align}
We note that the Banister model does not have bounds on the daily training impulse $w(i)$, which means that there is no upper bound on the level of performance $P(n)$ that can be achieved on day $n$. For the optimization in \eqref{eq:BanisterOptimizationProblem} to be a well-posed, bounded optimization problem, we require constraints on the maximum and minimum possible training impulse per day. The lower bound
\begin{align} \label{con:trainingconstraintLower}
0 \leq w(i), \text{ for all } i \in \{1, ... , n-1 \},
\end{align}
ensures that we cannot prescribe negative training. The upper bound that naturally arises from the model is
\begin{align} \label{con:trainingconstraintUpper}
k_2 w(i) \leq P_0 + \sum_{\ell=1}^{i-1} w(\ell)\left[  k_1 \exp{\left( - \frac{(i-\ell)}{\tau_1} \right)}- k_2 \exp{\left( - \frac{(i-\ell)}{\tau_2} \right)} \right] ~ \text{ for all } i \in \{1, ... , n-1 \}.
\end{align}
This upper bound ensures that the fatigue incurred from training, $k_2 w(i)$, cannot exceed the predicted performance on that day. Equivalently, because fatigue is expressed as a loss of performance, the model does not permit losing more performance than is available. Moreover, if a smaller upper bound $x$ were proposed, the model would still allow an additional training impulse of $P(i)/k_2 - x$, which would remain both feasible and performance-enhancing. Therefore, $x$ cannot be the true upper bound under the model.

\begin{figure}[ht!]
     \centering
         \begin{overpic}[width=0.4\textwidth,trim=50 0 0 0,clip]{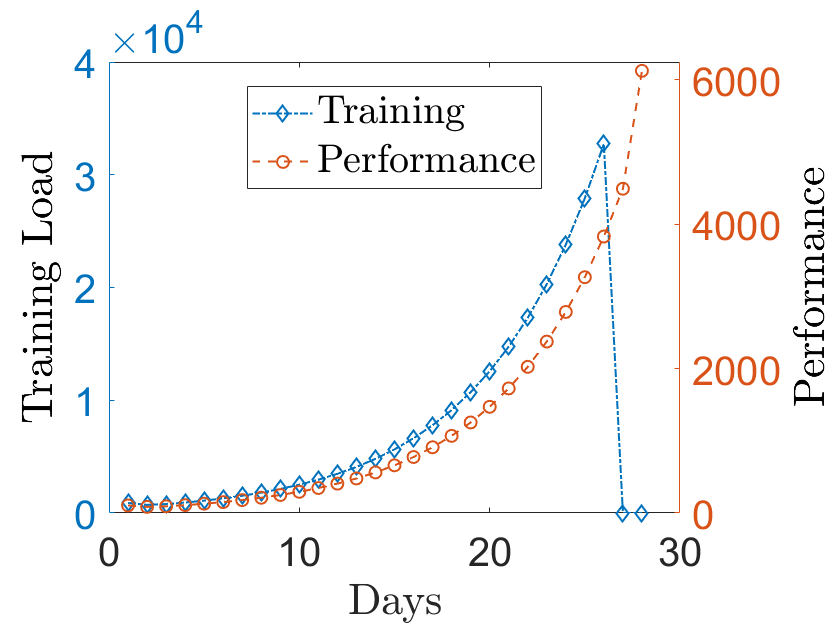}
            \put(-3,42){\rotatebox{90}{\makebox(0,0){Training impulse}}}
        \end{overpic}
        \caption{Optimized training impulse profile to maximize performance on day 28 using the Banister model with parameters: 
        $P_0      = 104.8,
    k_1      = 0.048,
    k_2      = 0.117,
    \tau_1    = 38,
    \tau_2    = 1.9$, taken from Table 3, mean values of 8 subjects, from \citep{busso1991adequacy}.
    For these parameters, $t_r = 1.78$ days and $t_p = 7.77$ days. }
        \label{fig:1BanisterModel}
\end{figure}

Since both the Banister model and the associated constraints are linear in the training impulse profile $\bm{w}$, the optimization problem in \eqref{eq:BanisterOptimizationProblem}--\eqref{con:trainingconstraintUpper} can be solved using linear programming. We did this using the \texttt{linprog} routine in \textsc{MATLAB}. The model parameter values were taken from \citep{busso1991adequacy}, in which eight participants completed a fourteen-week cycloergometer training program and the Banister model parameters were estimated from the resulting training impulse and performance data. Using the mean parameter values reported in Table~3 of \citep{busso1991adequacy}, we computed the training impulse profile that maximizes performance improvement over 28 days; the resulting profile is shown in Fig.~\ref{fig:1BanisterModel}.

In the experiment, the participants experienced an average improvement of 33.3~PU (Performance Units), increasing from 104.8~PU to 138.1~PU over the fourteen-week training period. In contrast, the optimization predicts that, under the training impulse profile shown in Fig.~\ref{fig:1BanisterModel}, an improvement of 6020.2~PU---from 104.8~PU to 6125~PU---should have been achievable in only four weeks. Furthermore, if the same optimal policy were maintained over the full fourteen weeks, the model predicts an improvement of $4.159 \times 10^8$~PU. Thus, the performance improvement predicted by the optimization exceeds that observed in practice by seven orders of magnitude. This discrepancy indicates that, without the inclusion of further physiological mechanisms, the Banister model produces predictions that are not physically plausible.

The optimal solution observed in Fig.~\ref{fig:1BanisterModel}, is not a pathological case. 
\begin{theorem} \label{theorem1}
The training impulse profile that solves the optimization problem in \eqref{eq:BanisterOptimizationProblem}, subject to the constraints in \eqref{con:trainingconstraintLower}--\eqref{con:trainingconstraintUpper}, can be derived recursively via the expressions
    \begin{align} \label{eq:TrainAsHardASPossible}
    w^*(i) =
    \begin{cases}
\frac{P(i)}{k_2} = \frac{1}{k_2} \left[ P_0 + \sum_{\ell=1}^{i-1} w^*(\ell) \left( k_1 \exp{ \left(- \frac{(i-\ell)}{\tau_1} \right) }- k_2 \exp{ \left(- \frac{(i-\ell)}{\tau_2} \right) } \right) \right], & \text{for all }\ i < n- t_r \\
0, & \text{for all } \  i \geq n- t_r 
\end{cases}
\end{align}
for all $0<k_1<k_2$ and $0< \tau_2 <\tau_1$. 
\end{theorem}
\noindent\textbf{Proof.}  See \ref{sec: Appendix}. 

Equation \eqref{eq:TrainAsHardASPossible} prescribes that we should train to our maximum allowable amount on all days $i$ where $i < n-t_r$ and then cease all training until the day of the performance. Examining its qualitative behavior and inspecting it for realistic elements, such as periodization and tapering, the Banister model will never produce periodization, nor tapering dynamics beyond the trivial taper of ceasing all training after $n-\trp$ days. This tapering style is not typically reported in accounts of elite athlete training behavior \citep[see, for example,][]{stellingwerff2012case}. Theorem~\ref{theorem1} is a surprising result relevant to the Banister model, and is not a generally true statement regarding all bounded linear impulse response functions with linear constraints -- see \ref{sec: AppendixA2} for several counterexamples of such optimization problems producing periodization. 


\begin{remarkunnumbered}[General upper-bound constraints]
The proof presented in \ref{sec: Appendix} is written for the specific upper-bound constraint
\eqref{con:trainingconstraintUpper}. However, the argument extends directly to any constraint of the form
\[
    C\,w(i) \le P(i),
\]
where $C$ is an arbitrary constant satisfying $C \in [k_2 - k_1, \infty)$. For example, one may conservatively enforce that fatigue constitute at most $10\%$ of the predicted performance by taking $C = 10\,k_2$, or impose that no residual performance remains once performance gains are accounted for by taking $C = k_2 - k_1$. In all such cases, the optimal solution retains the same structural form as in
\eqref{eq:TrainAsHardASPossible}; the only modification is the replacement $P(i)/k_2 \rightarrow P(i)/C$ throughout the analysis.
\end{remarkunnumbered}

\begin{remarkunnumbered}[Infinite-horizon equilibrium]
For $i < n-t_r$, Theorem~\ref{theorem1} implies that the optimal policy satisfies the equality
\[
    k_2 w^*(i)
    = P_0 + \sum_{\ell=1}^{i-1} w^*(\ell)\!
    \left(
        k_1 \exp\!\left(-\frac{i-\ell}{\tau_1}\right)
        - k_2 \exp\!\left(-\frac{i-\ell}{\tau_2}\right)
    \right).
\]
In the infinite-horizon setting $n\to\infty$, it is natural to consider constant equilibria of this recursion. Assuming this exists, setting $w^*(i)\equiv w_\infty$ and passing to the limit $i\to\infty$ yields the equation
\begin{align}
    w_{\infty}
    &=
    \frac{1}{k_2}\left[
        P_0
        + \sum_{\ell=1}^{\infty} w_{\infty}
        \left(
            k_1 \exp\!\left(-\frac{\ell}{\tau_1}\right)
            - k_2 \exp\!\left(-\frac{\ell}{\tau_2}\right)
        \right)
    \right] \nonumber \\
    &=
    \frac{P_0}{
        k_2
        - \dfrac{k_1}{\exp(1/\tau_1)-1}
        + \dfrac{k_2}{\exp(1/\tau_2)-1}
    },
    \label{eq:Training Asymptote}
\end{align}
where we used the identity
\[
    \sum_{\ell=1}^{\infty} \exp\!\left(-\frac{\ell}{\tau}\right)
    = \frac{1}{\exp(1/\tau)-1}.
\]
Provided the denominator in \eqref{eq:Training Asymptote} is positive, the equilibrium $w_\infty$ is finite and positive, equivalently
\begin{align}\label{eq:BanisterParameterConstrain}
    k_2\!\left(
        1 + \frac{1}{\exp\,(1/\tau_2)-1}
    \right)
    - \frac{k_1}{\exp\,(1/\tau_1)-1}
    > 0.
\end{align}
If the condition in \eqref{eq:BanisterParameterConstrain} is not satisfied, then it is not possible for the optimal training impulse profile during the phase when $i < n-t_r$ to converge to a constant finite value as $n \to \infty$. Consequently, if human performance is regarded as bounded, \eqref{eq:BanisterParameterConstrain} provides a natural admissibility condition when estimating Banister-model parameters.
\end{remarkunnumbered}


It is worth noting that if we are in the situation $O(n) \gg O(\tau_1)$ and the parameters satisfy the condition \eqref{eq:BanisterParameterConstrain}, early training will not contribute significantly to the final result. If we are in the unbounded case, however, early training will not persist to the final date, but will facilitate greater training in the in-between time and will indirectly lead to an increased final result.

In conclusion, optimization of the Banister model recommends that we train maximally to day $n-\trp$, then not at all. 
To induce more complex behavior in the optimized training impulse profile that is more consistent with elite athlete training behavior noted in the sports science literature, such as periodization and nuanced tapering strategies, alterations to the model are necessary. 



\subsection*{The Busso Model}

Returning to the question of ``What training impulse profile should an athlete adopt to maximize their performance on day $n$?'' Let us now answer this question in the case of the Busso model. Mathematically speaking, this translates to solving the optimization problem
\begin{align} \label{eq:BussoOptimizationProblem}
    \max_{\bm{w}} ~ P_0 + \sum_{i=1}^{n-1} w(i)\left[ k_1 \exp{\left( - \frac{(n-i)}{\tau_1} \right)}- \sum_{j=1}^{i-1} k_3 w(j)  \exp{\left( - \frac{(i-j)}{\tau_3} \right)}\exp{\left( - \frac{(n-i)}{\tau_2} \right)} \right].
\end{align}
As with the Banister model, unless we introduce constraints on the allowable daily training impulse, the optimization problem in \eqref{eq:BussoOptimizationProblem} is unbounded. Further, when we seek to optimize the Busso model with constraints derived using the same logic as \eqref{con:trainingconstraintLower}--\eqref{con:trainingconstraintUpper}, we arrive at the lower bound
\begin{align} \label{con:trainingconstraintsBusso_11}
0 \leq w(i), \text{ for all } i \in \{1,2,\dots,n-1\},
\end{align}
and upper bound
\begin{align} \label{con:trainingconstraintsBusso_12}
\sum_{j=1}^{i-1} w(j) k_3 \exp{\left( - \frac{(i-j)}{\tau_3} \right)} w(i) \leq P_0 + \sum_{\ell=1}^{i-1} w(\ell)\left[ k_1 e^{\left( - \frac{(i-\ell)}{\tau_1} \right)}- \sum_{j=1}^{\ell-1} k_3 w(j)  e^{\left( - \frac{(\ell-j)}{\tau_3} \right)} e^{\left( - \frac{(i-\ell)}{\tau_2} \right)} \right],
\end{align}
for all $ i \in \{1,2, ... , n-1 \}$. Even with this upper bound, we still encounter a nonphysical outcome: that infinite training, and hence infinite performance improvement, is possible. This is due to the fact that the fatigue magnitude, $k_2(i)$, varies depending on the recent history of training and can be made arbitrarily small by simply waiting long enough after the last training impulse. Then, since the fatigue magnitude can be made arbitrarily small, we can theoretically have an arbitrarily large training impulse which produces an arbitrarily large performance improvement which is realized from the subsequent day. This nonphysical behavior largely precludes the Busso model from being a practical optimization tool. 

Supposing we wanted to force the Busso model to form a well-posed, bounded optimization problem, we can, out of necessity, add an arbitrary upper bound to the training impulse profile
\begin{align} \label{con:trainingconstraintsBusso_2}
w(i) \leq w_U ~ \text{ for all } i \in \{1,2, ... , n-1 \}.
\end{align}
We note that it would be poor mathematical modeling if an optimal solution is significantly influenced by an arbitrary decision made in the modeling process. In this case, we expect that optimization of the Busso model will be significantly impacted by the arbitrary upper bound on the training impulse profile. Perhaps the only nontrivial recommendation might be how often we train to this arbitrary upper bound, and what the expected final performance outcome is. 

The optimization problem in \eqref{eq:BussoOptimizationProblem}, subject to the constraints in \eqref{con:trainingconstraintsBusso_11}--\eqref{con:trainingconstraintsBusso_2}, constitutes a quadratically-constrained optimization problem with a quadratic objective function. It unfortunately does not, however, satisfy the negative semidefinite condition to guarantee a concave optimization landscape. We therefore treated the optimization problem as we would a more general constrained, nonlinear optimization problem. We solved the Busso optimization problems using \textsc{MATLAB}'s inbuilt \texttt{fmincon}. This is a gradient-based algorithm that finds a local optimum. By using a variety of starting values, an estimate of the global optimum can be obtained by selecting the maximum from the resulting set of local optima. We can also assess our confidence in our estimate of the global optimum by examining characteristics of the set of local optima; for example, a consistent outcome, or a tightly clustered outcome, suggests that the maximum is robust to initial conditions and may be close to the true global optimum.

\begin{figure}[ht!]
     \centering
     \begin{subfigure}{0.4\textwidth}
         \centering
         \begin{overpic}[width=0.8\textwidth,trim=40 0 0 0,clip]{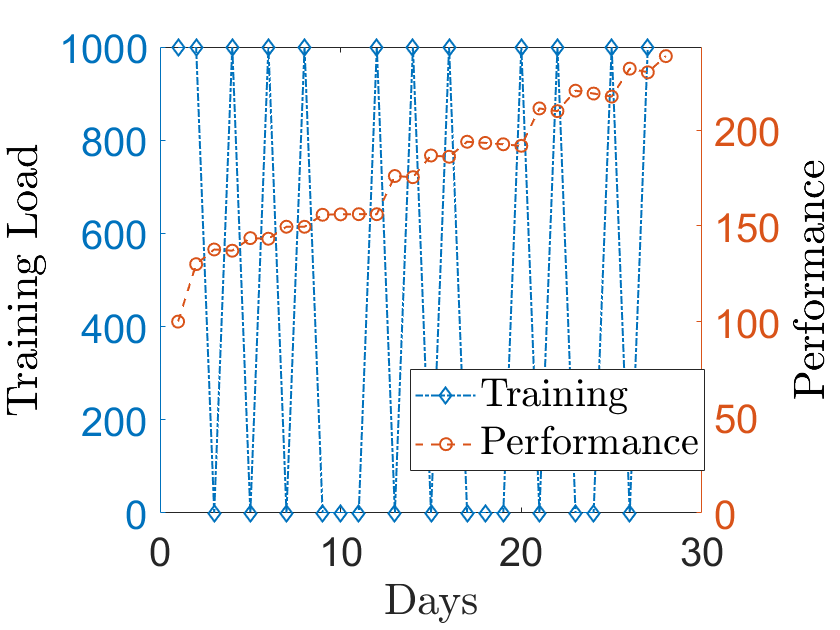}
         \put(-2,42){\rotatebox{90}{\makebox(0,0){\small Training impulse}}}
        \end{overpic}
         \caption{ }
         \label{subfig:g10}
     \end{subfigure}
     \begin{subfigure}{0.4\textwidth}
         \centering
         \begin{overpic}[width=0.8\textwidth]{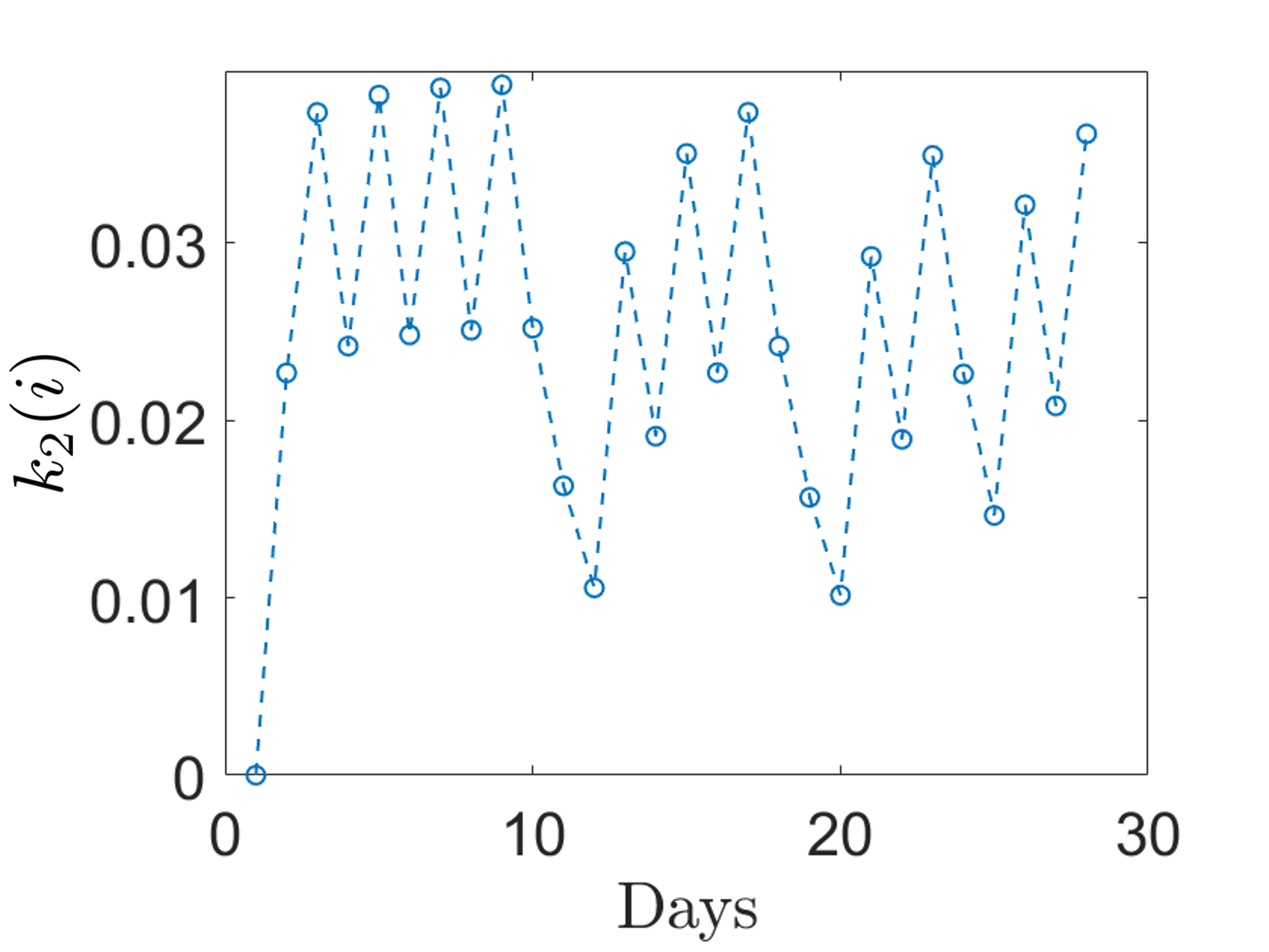}
         \end{overpic}
         \caption{ }
         \label{subfig:g11}
     \end{subfigure}
     \centering
     \begin{subfigure}{0.4\textwidth}
         \centering
         \begin{overpic}[width=0.8\textwidth,trim=40 0 0 0,clip]{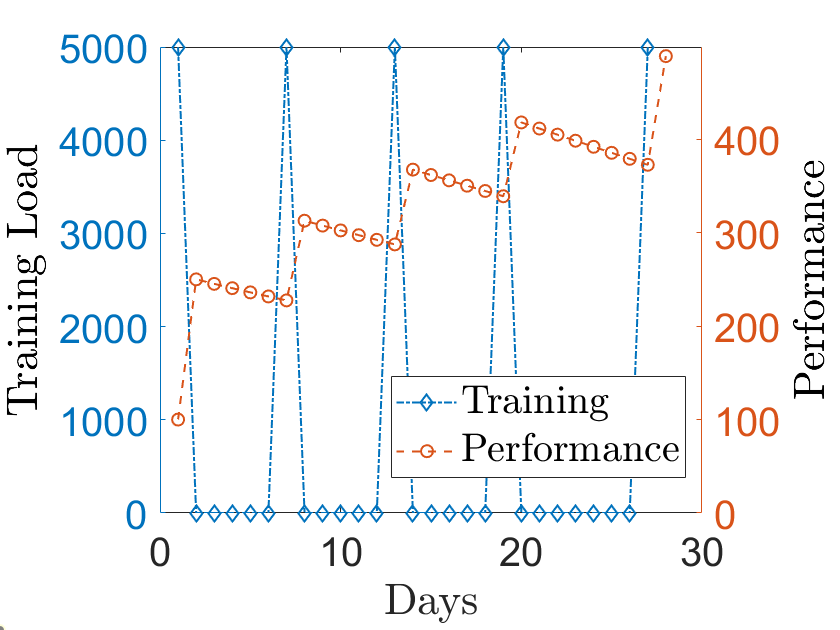}
         \put(-3,42){\rotatebox{90}{\makebox(0,0){\small Training impulse}}}
        \end{overpic}
         \caption{ }
         \label{subfig:g14}
     \end{subfigure}
     \begin{subfigure}{0.4\textwidth}
         \centering
         \begin{overpic}[width=0.8\textwidth]{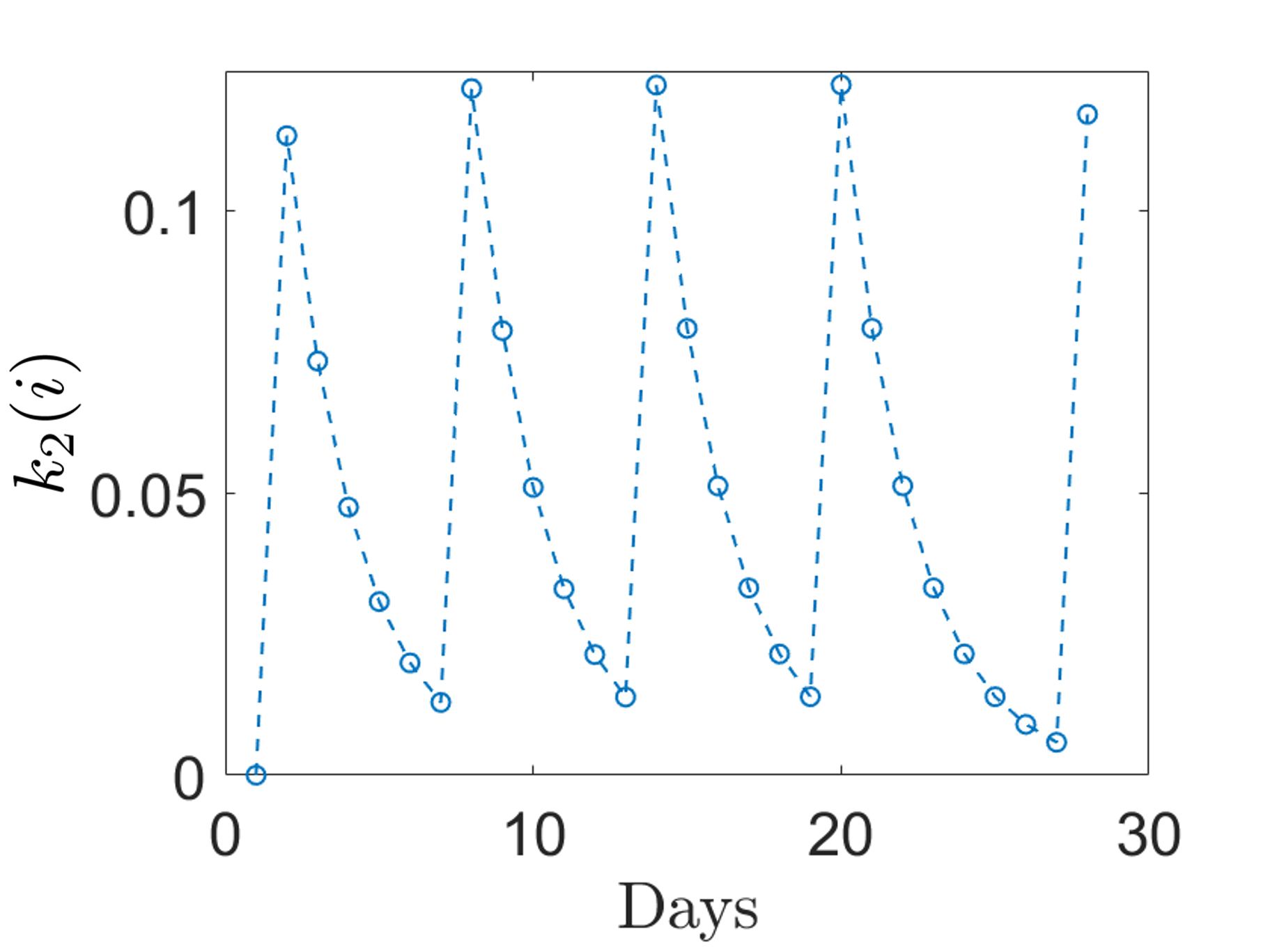}
         \end{overpic}
         \caption{ }
         \label{subfig:g15}
     \end{subfigure}
        \caption{(a) Optimized training impulse profile to maximize performance on day 28 using the Busso model, with parameters taken from Table 2 of \citep{busso2003variable}: 
        $P_0      = 100,
    k_1      = 0.031,
    k_3      = 0.000\,035,
    \tau_1    = 30.8,
    \tau_2    = 16.8,
    \tau_3    = 2.3$ and with an arbitrary upper-bound of $w_U = 1\,000$ ATU.
    (b) The variation in fatigue cost $k_2(i)$ as a function of time, for (a). (c) Same as (a) but with an arbitrary upper bound of $w_U = 5\,000$ ATU. (d) Same as (b) but relevant to (c). }
        \label{fig:VDR}
\end{figure}

Two optimal training programs, which differ in the arbitrary training impulse upper bound, are observable in Fig.~\ref{fig:VDR} with parameters taken from Table 2 of \citet{busso2003variable}. Firstly, we can see that the solutions are dependent on the arbitrary upper-bound, with Fig.~\ref{subfig:g10}-\ref{subfig:g11} having an upper bound of $w_U = 1\,000$ and Fig.~\ref{subfig:g14}-\ref{subfig:g15} having an upper bound of $w_U = 5\,000$. Present in both is the ramping up and decaying away of the fatigue cost $k_2(i)$, which creates periodization of on-off days. This is due to the optimization algorithm suggesting that once the fatigue magnitude $k_2(i)$ exceeds the fitness magnitude $k_1 = 0.031$, then we should refrain from training. In the case that $k_2(i)<k_1$, then we also see an immediate improvement in performance on the subsequent day for a maximal training impulse, which does not correspond to the expected fatigue and improvement response following maximal training. In effect, this means that the optimization problem in \eqref{eq:BussoOptimizationProblem} is not meaningfully determined by the model but is dominantly driven by the interaction of the arbitrary upper bound on the daily training impulse $w_U$ and the fatigue magnitude decay time-constant $\tau_3$. Seemingly, whilst the Busso model has increased explanatory power over the Banister model \citep{busso2003variable}, it lacks practicality as an optimization tool.

\section{Proposed Modifications to the Fitness--Fatigue Model} \label{Sec:Model Modifications}

\citet{mujika2003scientific} identified three types of tapering behavior: linear, exponential, and step-down. A central aim of this section is to investigate whether these tapering structures can arise as optimal training profiles from fitness--fatigue models with suitable nonlinear modifications.

The Banister model, which is linear, predicted the optimal training impulse profile $\bm{w^*}$ in \eqref{eq:TrainAsHardASPossible} that maximizes performance on day $n$; we regard this as a baseline reference case for future comparison. Accordingly, the extent to which modified models generate features such as periodization or tapering should be evaluated relative to this canonical solution. The nonlinearity introduced in the Busso model largely defied optimization and produced nonphysical performance improvements. In neither of these models did we observe periodization and tapering in the forms recommended by empirical studies.

We propose a systematic exploration of nonlinear modifications to the Fitness--Fatigue Impulse Response model of the form
\begin{align} \label{eq:CombinedChanges}
    P(n) = P_0 + \sum_{i=1}^{n-1} M_{F}(w(i),f(i)) \exp{ \left( \frac{-(n-i)}{\tau_{1}(w(i))} \right)} - k_{2} w(i) \exp{ \left( \frac{-(n-i)}{\tau_{2}(w(i))} \right)},
\end{align} 
where $M_F (w(i),f(i)) $ is the fitness gain as a function of the training impulse $w(i)$ and cumulative fatigue $f(i)$ defined below equation \eqref{Eq:OGBanisterDiscrete}, and $\tau_1(w(i))$ and $\tau_2(w(i))$ are the impulse response decay constants as a function of training impulse. Specifically, we investigate these nonlinearities in the immediate fitness gain response to training, Subsection \ref{Sec:Magnitude Modifications}, in the time-decay of the impulse response, Subsection \ref{Sec:Time-Constant Based Alterations} and the combination of the two, Subsection \ref{Sec:Combined Changes}. A natural extension of \eqref{eq:CombinedChanges} would be to introduce analogous nonlinear modifications to the fatigue response. This would, however, amount to redefining the training impulse itself. The definition and calibration of such a quantity is a substantial topic in its own right and is beyond the scope of the present study. Our goal is to identify nonlinearities that, when we solve
\begin{align} \label{CombinedOptimizationProblem}
    \max_{\bm{w}} ~ P_0 + \sum_{i=1}^{n-1} M_{F}(w(i),f(i)) \exp{ \left( \frac{-(n-i)}{\tau_{1}(w(i))} \right)} - k_{2} w(i) \exp{ \left( \frac{-(n-i)}{\tau_{2}(w(i))} \right)},
\end{align}
subject to 
\begin{align*}
    0 \le k_2w(i) \le P_0 + \sum_{i=1}^{n-1} M_{F}(w(i),f(i)) \exp{ \left( \frac{-(n-i)}{\tau_{1}(w(i))} \right)} - k_{2} w(i) \exp{ \left( \frac{-(n-i)}{\tau_{2}(w(i))} \right)}
\end{align*}
for all $i \in \{1, \dots , n-1\}$, we produce optimized training impulse profiles that more qualitatively align with empirically derived guidelines for patterns such as periodization and tapering. The optimization problem in \eqref{CombinedOptimizationProblem} is a constrained, nonlinear optimization problem. All versions of the optimization problem were solved using \textsc{MATLAB}’s built-in \texttt{fmincon} function. In general, the algorithm found the problem to be tractable; however, when periodization was present, the optimization landscape exhibited multiple local optima. To approximate the global optimum in these cases, we employed multiple initializations of the algorithm and selected the maximum performance outcome among the resulting local optima.


\subsection{Magnitude Modifications} \label{Sec:Magnitude Modifications}

The Banister model employs the assumption that each unit of training produces a fitness gain of $k_1$. The logical next step is to consider formulations in which we allow the fitness gain to vary with the training impulse or context of the training impulse. It could be that as a training session goes on, there are diminishing returns with increasing training impulse. Or more broadly speaking, it could be that as fatigue is accumulated, there are diminished returns for the same training impulse. 

\begin{figure}[ht!]
     \centering
     \begin{subfigure}{0.4\textwidth}
         \centering
         \begin{overpic}[width=0.8\textwidth]{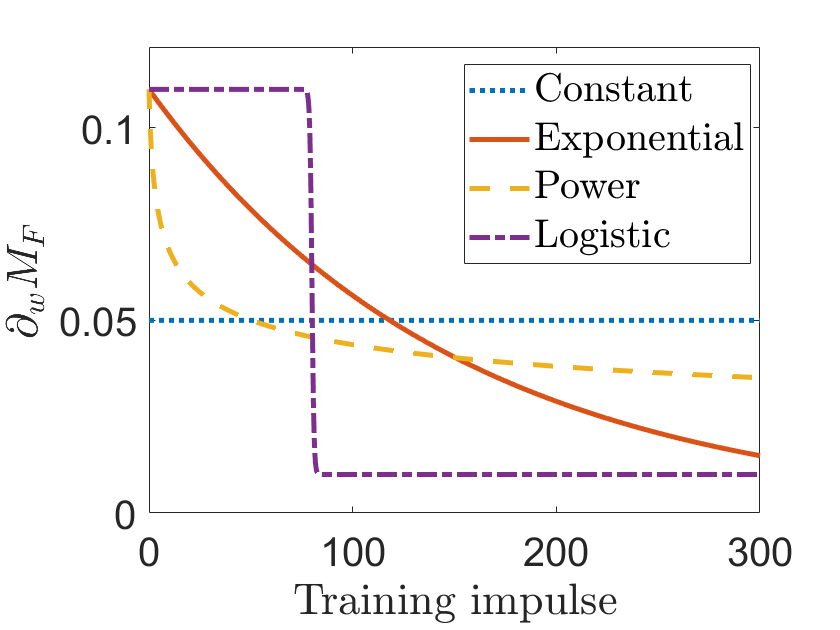}
        \end{overpic}
         \caption{ }
         \label{subfig:gr21}
     \end{subfigure}
     \begin{subfigure}{0.4\textwidth}
         \centering
         \begin{overpic}[width=0.8\textwidth]{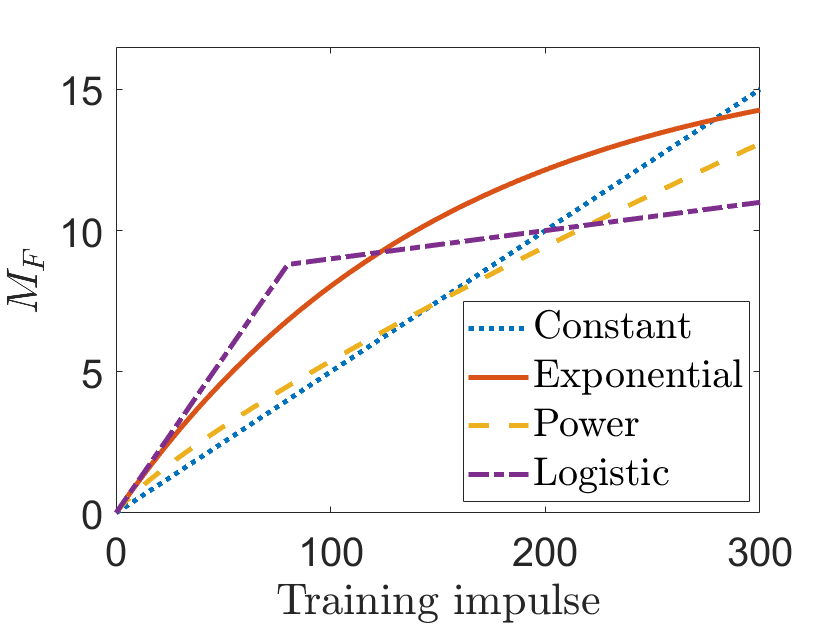}
         \end{overpic}
         \caption{ }
         \label{subfig:gr22}
     \end{subfigure}
     \centering
\caption{
(a) Illustrative fitness gain function derivatives, $\partial_w M_F(w)$ and (b) fitness gain functions $M_F$, for the constant (Banister) case $k_1 = 0.05$, exponential decay \eqref{eq:MMexponential} with $k_1 = 0.11$, $k_3 = 150$,
power decay \eqref{eq:MMpower} with $k_1 = 0.11$, $k_3 = 0.2$, 
and logistic decay \eqref{eq:MMlogistic} with $k_1 = 0.01$, $k_3 = 2$, $k_4 = 0.1$, $k_5 = 80$.
}
        \label{fig:M_F}
\end{figure}

To capture this idea, we can consider a number of fitness gain function modifications to the current linear form. For the sake of simplicity, we will consider the perspective of the derivative of the fitness gain function. For example, the Banister model has a fitness gain function that is linear in $w(i)$ and the derivative is a constant. Here, we will consider the fitness gain function derivatives:
\begin{enumerate}
    \item Exponential decay
    \begin{align} \label{eq:MMexponential}
        M_F(w(i)) = \int_{0}^{w(i)} k_1 \exp{ \left( -\frac{u}{k_3} \right)} du = k_1 k_3 \left( 1-\exp{ \left(-\frac{w(i)}{k_3} \right)} \right),
    \end{align}
    \item Power decay
    \begin{align} \label{eq:MMpower}
        M_F(w(i)) = \int_{0}^{w(i)} \frac{k_1}{(1+u)^{k_3}} du = \frac{k_1}{k_3-1} \left( 1- \frac{1}{(1+w(i))^{k_3-1}} \right),
    \end{align}
    \item Logistic decay
    \begin{align}\label{eq:MMlogistic}
        M_F(w(i)) = \int_{0}^{w(i)} k_1 + \frac{k_4}{1 + \exp{ (k_3(u-k_5)} )} du = k_1 w(i) +  \frac{k_4}{k_3} \ln \left( \frac{1 + \exp{(k_3 k_5)}}{1 + \exp{(k_3[k_5 - w(i)])}} \right),
    \end{align}
\end{enumerate}
where $k_3 > 0$ is a parameter that adjusts the decay of fitness gain function, and $k_4 >0$ and $k_5>0$ occur in the logistic case, which controls the increase for relatively small training impulses and the offset location, respectively. An comparative illustration of the different cases of diminishing returns and fitness gain functions can be seen in Fig.~\ref{fig:M_F}. Note, nonlinear fitness gain functions have previously been considered in the athletic performance modeling literature (see  \citet{hellard2005modeling}, although, their focus was on fitting models with saturation thresholds, rather than the optimization behavior).

In the Banister model, there was a single time to return, $t_r$, and a single time to the peak of a training impulse, $t_p$, defined in \eqref{eq:Time-to-return} and \eqref{eq:Time-to-peak}, respectively. With diminishing returns, however, these points are now curves. For each $w(i)$, there will now be a unique $t_r$ and $t_p$, and for each $t_p$, there will be a corresponding peak value of the impulse response. 

For an analytical example, when the derivative of the fitness gain function is a decaying exponential, we can calculate the time to return as
\begin{align}
    t_r = \frac{\tau_1 \tau_2}{\tau_1 - \tau_2} \ln \left( \frac{k_2 w(i)}{k_1k_3 \left( 1-\exp{(-w(i)/k_3)} \right)} \right).
\end{align}
Similarly, the time to a peak in training effect will be
\begin{align}
    t_p = \frac{\tau_1 \tau_2}{\tau_1 - \tau_2} \ln \left( \frac{k_2 \tau_1 w(i)}{k_1k_3 \tau_2 \left( 1-\exp{(-w(i)/k_3)} \right)} \right).
\end{align}

By having a variable time to return on training benefit, this should allow for more complex tapering dynamics to arise naturally from the model. Previously, we obtained that optimization of the training impulse profile to maximize the performance from the Banister model produced $\bm{w^*}$ in \eqref{eq:TrainAsHardASPossible}, which had the tapering strategy of complete rest after day $n - \trp $. In practice, however, we do not typically see this type of tapering behavior utilized by elite athletes \citep[see, for example,][]{stellingwerff2012case}. In previous studies of tapering in Fitness--Fatigue models \citep{banister1999training,mujika2003scientific,thomas2009computer}, authors considered a set of realistic tapering strategies and compared Fitness--Fatigue model's prediction for performance outcomes. Some strategies proposed by authors included: a step-down in consistent training, a linear decrease in training and an exponential decrease in training with different decay rates, as well as the concept of an ``over-training'' block preceding the taper block. With modifications in the magnitude representation, however, we no longer have to supply the model with a set of realistic options. Instead, as we will see, the model naturally gives rise to these options depending on the form of the diminishing returns.

It is possible to analytically calculate the expression for the optimum training impulse taper for each case of fitness magnitude modification \eqref{eq:MMexponential}--\eqref{eq:MMlogistic}. If we assume that the training impulse is below the upper training constraint in \eqref{CombinedOptimizationProblem}, such as during the taper, then the days are independent---which is to say that it is synonymous to optimize each day individually, as it is to optimize the days collectively in series.

In the Banister model, optimizing a single day simply led to the conclusion that more was better. That, if we are $j$ days from our desired maximum performance date, the return on training was
\begin{align}
    w(j)\left[ k_1 \exp{\left( - \frac{j}{\tau_1} \right)}- k_2 \exp{\left( - \frac{j}{\tau_2} \right)} \right]
\end{align}
and so long as $j>t_r$, the greater $w(j)$, the greater the return. Put another way, there is no point at which increasing $w(j)$ yields less performance on the prioritized day. In the modified magnitude case, however, we now see a point at which $\partial P(n)/\partial w(i) < 0$, and we can solve for the optimum amount of training. In the exponential decay case in equation \eqref{eq:MMexponential}, we see that
\begin{align}
  0 &= \frac{d}{dw} \left( k_1 k_3 \left( 1-\exp{ \left(-\frac{w}{k_3} \right)} \right) \exp{\left( -\frac{j}{\tau_1} \right)} - k_2 w \exp{ \left(-\frac{j}{\tau_2} \right)} \right) \nonumber \\
    \Rightarrow w(j) &= k_3 \ln \left( \frac{k_1}{k_2} \right) + j k_3 \left( \frac{\tau_1 - \tau_2}{\tau_1 \tau_2} \right).
    \label{eq:linear-taper}
\end{align}
This tells us that as $j$ counts down to our prioritized performance date, we should linearly decrease our training impulse profile to obtain an optimum result. The extent of this taper is determined by the intersection of the taper and maximum training constraint, $k_2w(i) \leq P(i)$. In summary, this indicates that an exponential decay in the derivative of the fitness gain function naturally gives rise to a linear taper towards the desired maximum performance date. 

Performing the same calculation for a power decay as in equation \eqref{eq:MMpower}, we find that the optimal training impulse as we approach the prioritized maximum performance date is given by
\begin{align}
  0 &= \frac{d}{dw} \left( \frac{k_1}{k_3-1} \left( 1- \frac{1}{(1+w)^{k_3-1}} \right) \exp{ \left( -\frac{j}{\tau_1} \right)} - k_2 w \exp{ \left( -\frac{j}{\tau_2} \right)} \right) \nonumber \\
    \Rightarrow w(j) &=  \left( \frac{k_1}{k_2} \right)^{\frac{1}{k_3}} \exp{ \left( \frac{j}{k_3} \left( \frac{\tau_1 - \tau_2}{\tau_1 \tau_2} \right) \right) } - 1.
    \label{eq:exponential-taper}
\end{align}
This tells us that as $j$ counts down to the prioritized performance date, we should exponentially taper our training impulse profile. Finally, we can perform the same calculation for a logistic decay as per equation \eqref{eq:MMlogistic}, for which we obtain
\begin{align}
      0 &= \frac{d}{dw} \left( \left[ k_1 w +  \frac{k_4}{k_3} \ln \left( \frac{1 + \exp{(k_3 k_5)}}{1 + \exp{(k_3(k_5 - w))}} \right) \right] \exp{ \left( -\frac{j}{\tau_1} \right)} - k_2 w \exp{ \left( -\frac{j}{\tau_2} \right)} \right) \nonumber \\
    \Rightarrow w(j) &=  k_5 + \frac{1}{k_3} \ln \left( \frac{k_4}{k_2 \exp{ \left(-j \left(\frac{\tau_1 - \tau_2}{\tau_1 \tau_2} \right) \right)} - k_1} - 1 \right).
    \label{eq:step-taper}
\end{align}
This is less analytically transparent compared to the exponential and power decay cases. For small $j$, this tells us to train at around the logistic offset value, $k_5$, and, depending on the value of $k_3$, to more or less maintain that value until the denominator of the term within the logarithm reaches zero, which occurs when $j = \tau_1 \tau_2/(\tau_1 - \tau_2) \ln(k_2/k_1)$. After this point in time, the analytical result no longer applies and we are again constrained by the upper limit of what is physically possible $k_2w(i) \leq P(i)$. Returning to the statement of ``more or less maintain [$k_5$]'' depending on $k_3$, if $k_3 \gg 1$ then this behaves much like a step down in training. If $k_3 \ll 1$ it will suggest a linear taper from the time $j = \tau_1 \tau_2/(\tau_1 - \tau_2) \ln(k_2/k_1)$. If $k_3 = O(1)$, then this will suggest a step-down taper which behaves approximately linear once stepped down. 

\begin{figure}[ht!]
     \centering
     \begin{subfigure}{0.32\textwidth}
         \centering
         \begin{overpic}[width=0.95\textwidth,trim=40 0 0 0,clip]{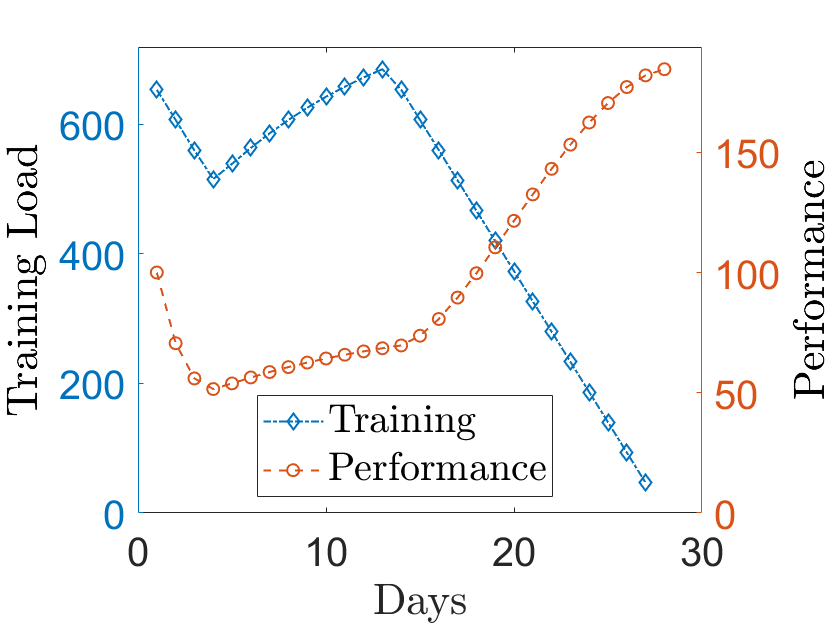}
         \put(-3,42){\rotatebox{90}{\makebox(0,0){\small Training impulse}}}
        \end{overpic}
         \caption{ }
         \label{subfig:g52}
     \end{subfigure}
     \begin{subfigure}{0.32\textwidth}
         \centering
         \begin{overpic}[width=0.95\textwidth,trim=40 0 0 0,clip]{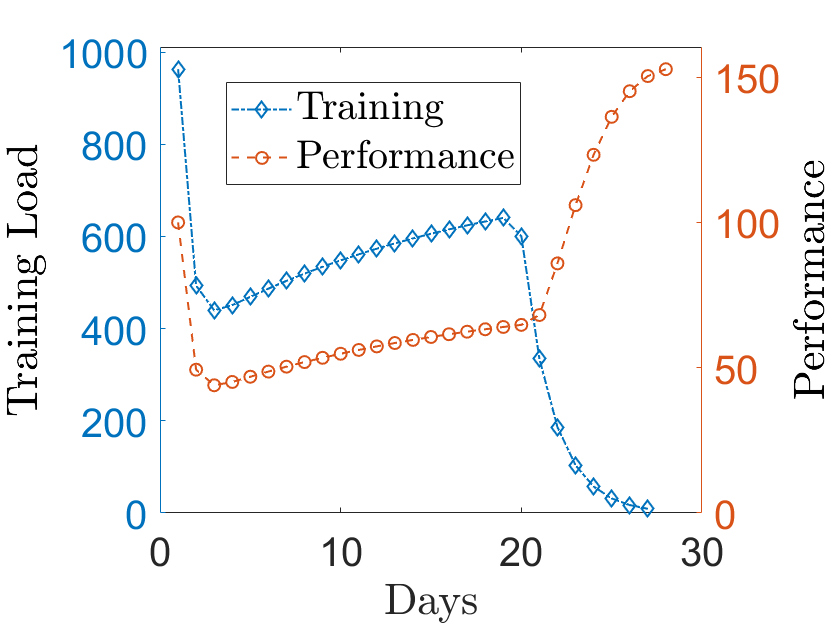}
         \put(-2,42){\rotatebox{90}{\makebox(0,0){\small Training impulse}}}
        \end{overpic}
         \caption{ }
         \label{subfig:g53}
     \end{subfigure}
     \begin{subfigure}{0.32\textwidth}
         \centering
         \begin{overpic}[width=0.95\textwidth,trim=40 0 0 0,clip]{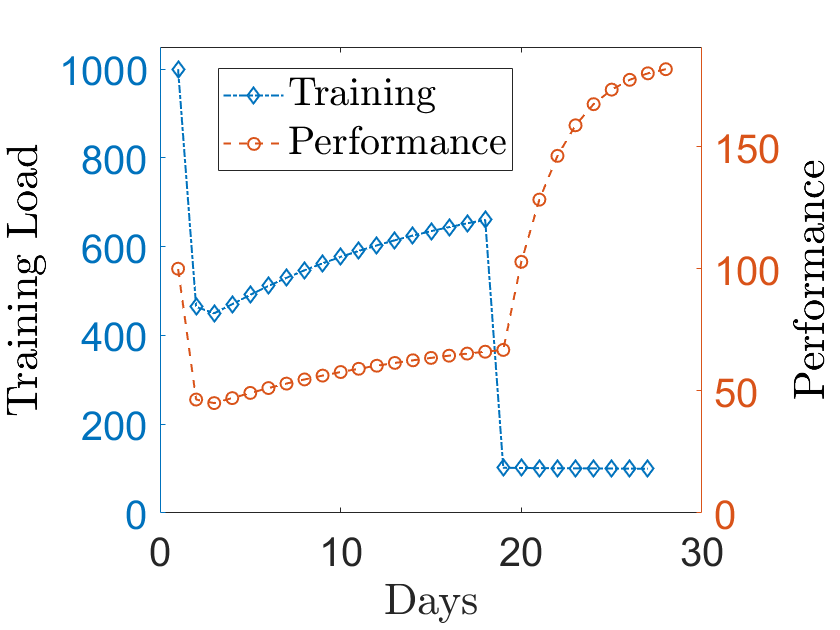}
         \put(-2,42){\rotatebox{90}{\makebox(0,0){\small Training impulse}}}
        \end{overpic}
         \caption{ }
         \label{subfig:g54}
     \end{subfigure}
        \caption{Optimized training impulse profiles to maximize performance on day 28 using the modified Fitness--Fatigue model with diminishing returns as captured by:
        (a) Exponential decay \eqref{eq:MMexponential} with $k_1 = 0.1$, $k_3 = 200$, 
        (b) Power decay \eqref{eq:MMpower} $k_1 = 0.2$, $k_3 = 0.4$, 
        (c) Logistic decay \eqref{eq:MMlogistic} $k_1 = 0.01$, $k_3 = 2$, $k_4 = 0.1$, $k_5 = 100$. 
        All with the parameters:
    $P_0  = 100,
    k_2   = 0.1,
    \tau_1 = 10,
    \tau_2 = 3$.}
        \label{fig:TaperingSolutions}
\end{figure}

For an example of the optimal training impulse profiles that maximize performance using the modified Fitness--Fatigue Impulse Response model, with exponential, power and logistic diminishing returns in fitness gain, see Fig.~\ref{fig:TaperingSolutions}. In this, we can see tapering strategies in the fashion that was analytically predicted: linear, exponential and step-down, as defined in  \eqref{eq:linear-taper}--\eqref{eq:step-taper}, for Fig.~\ref{subfig:g52}, \ref{subfig:g53} and \ref{subfig:g54}, respectively. 

\begin{figure}[ht!]
     \centering
     \begin{subfigure}{0.32\textwidth}
         \centering
         \begin{overpic}[width=0.95\textwidth,trim=40 0 0 0,clip]{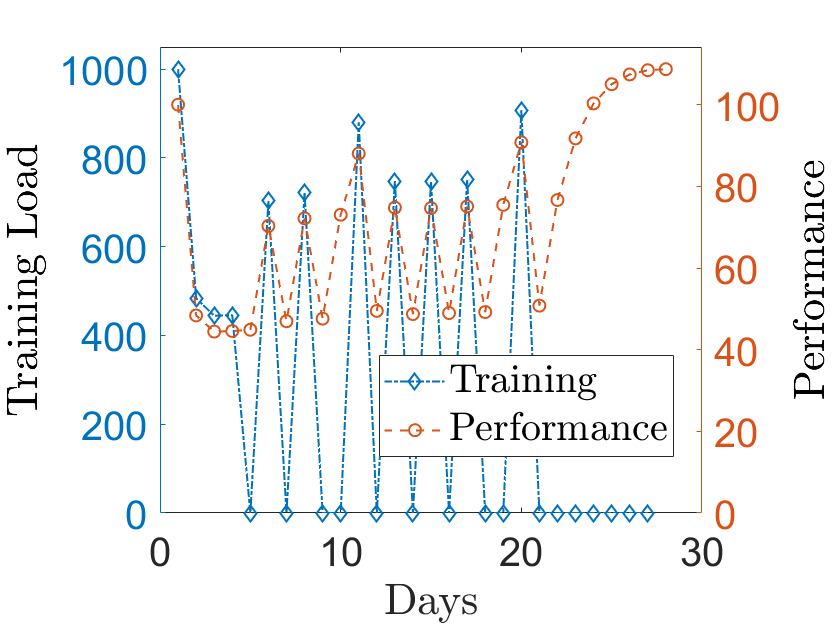}
         \put(-2,42){\rotatebox{90}{\makebox(0,0){\small Training impulse}}}
        \end{overpic}
         \caption{ }
         \label{subfig:g55_1}
     \end{subfigure}
     \begin{subfigure}{0.32\textwidth}
         \centering
         \begin{overpic}[width=0.95\textwidth,trim=40 0 0 0,clip]{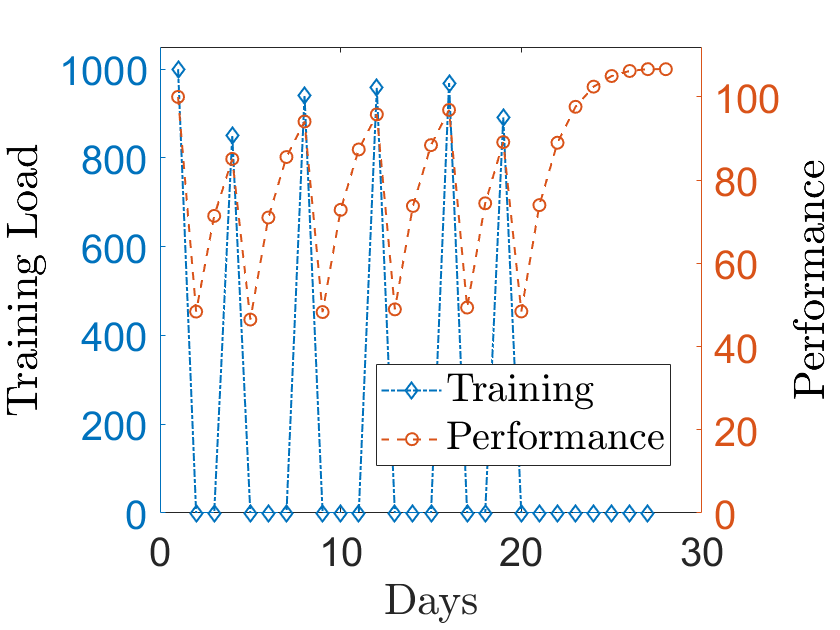}
         \put(-2,42){\rotatebox{90}{\makebox(0,0){\small Training impulse}}}
        \end{overpic}
         \caption{ }
         \label{subfig:g55_2}
     \end{subfigure}
     \begin{subfigure}{0.32\textwidth}
         \centering
         \begin{overpic}[width=0.95\textwidth,trim=40 0 0 0,clip]{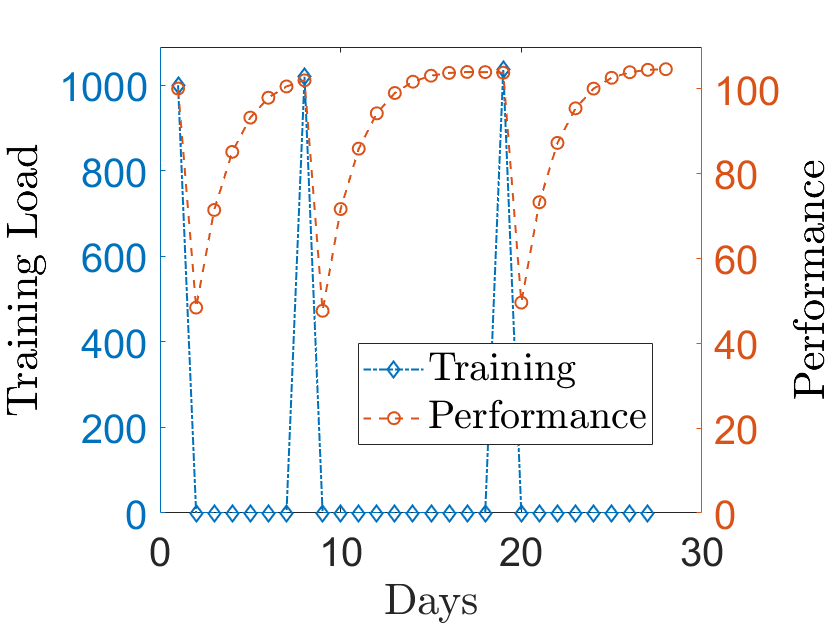}
         \put(-2,42){\rotatebox{90}{\makebox(0,0){\small Training impulse}}}
        \end{overpic}
         \caption{ }
         \label{subfig:g55_3}
     \end{subfigure}
        \caption{Optimized training impulse profiles to maximize performance on day 28 using the modified Fitness--Fatigue model with nonlinear fatigue feedback \eqref{eq:fatiguefeedback1} for the parameters:
        (a) $k_4 =100$, (b) $k_4 = 50$, 
        (c) $k_4 = 10$ and all other parameters were:
    $P_0  = 100,
    k_1   = 0.01,
    k_2   = 0.1,
    \tau_1 = 10,
    \tau_2 = 2$.
    }
        \label{fig:periodizationSolutions}
\end{figure}

Whilst we have seen that diminishing returns induces tapering dynamics in the optimized training impulse, it does not seem to induce periodization. To extend the idea of diminishing returns within a single training impulse to cover several training impulses, we will include the fatigue accumulated from previous training impulses as a decaying effect on the current training impulse's fitness gain in an effect we will call fatigue feedback. In what follows, we distinguish between fatigue feedback acting alone and fatigue feedback acting in combination with diminishing returns. Mathematically speaking, this could be represented in numerous ways. Selecting a couple for the point of demonstration, fatigue feedback could be captured by: 
\begin{enumerate}
    \item Fatigue feedback without diminishing returns, via an exponential decay,
    \begin{align} \label{eq:fatiguefeedback1}
        M_F(w(i),f(i)) = k_1 w(i) \exp{ \left(-\frac{f(i)}{k_4} \right)},
    \end{align}
    where $k_4$ is a constant that controls the extent of the deleterious effect of fatigue. Note, whilst we have just argued for the inclusion of diminishing returns as per equations \eqref{eq:MMexponential}-\eqref{eq:MMlogistic}, we have taken these out above to explore the effect of fatigue feedback in isolation. 
    \item Fatigue feedback with diminishing returns, such as having a power decay as in equation \eqref{eq:MMpower} and an exponential decay for fatigue sustained from previous days,
    \begin{align} \label{eq:doubleampdecay}
        M_F(w(i),f(i)) = \frac{k_1}{k_3-1}   \left( 1- \frac{1}{(1+w(i))^{k_3-1}} \right) \exp{ \left( -\frac{f(i)}{k_4} \right)}.
    \end{align}
\end{enumerate}
Although the fatigue feedback is expressed as an exponential decay in \eqref{eq:fatiguefeedback1} and \eqref{eq:doubleampdecay}, it could also be a power decay, logistic decay, or some other decaying function. Moreover, it could be combined with any of the same day effects modeled by exponential, power or logistic decays, as defined in \eqref{eq:MMexponential}--\eqref{eq:MMlogistic}. Numerically examining the first case of allowing fatigue feedback expressed in \eqref{eq:fatiguefeedback1}, on the modified Fitness--Fatigue model in \eqref{eq:CombinedChanges}, we can see that the optimal training impulse profile to maximize performance in 28 days has changed from training every day, to training only on select days---compare Fig.~\ref{fig:1BanisterModel} to Fig.~\ref{fig:periodizationSolutions}. This could be one mechanism underlying the training practice of periodization. Indeed, if the fitness gain function has a relatively high sensitivity to fatigue (reflected in a small value of $k_4$) then we see large rest times occurring between hard training days, Fig.~\ref{subfig:g55_3}. As the parameter $k_4$ is increased, however, we see the rest time between training impulses reduce and the amount of sustained fatigue increase, Fig.~\ref{subfig:g55_1}.

\begin{figure}[ht!]
     \centering
     \begin{subfigure}{0.32\textwidth}
         \centering
         \begin{overpic}[width=0.95\textwidth,trim=40 0 0 0,clip]{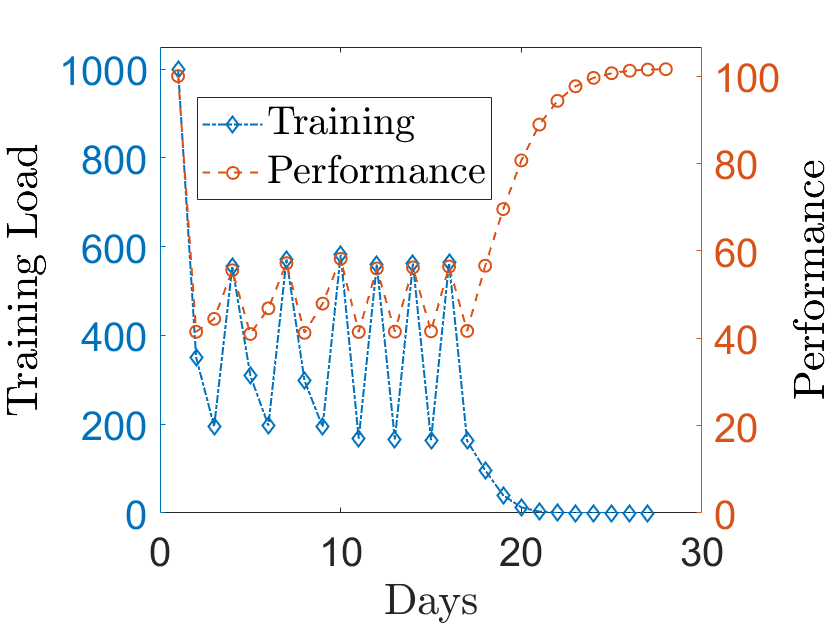}
         \put(-2,42){\rotatebox{90}{\makebox(0,0){\small Training impulse}}}
        \end{overpic}
         \caption{ }
         \label{subfig:g59_1}
     \end{subfigure}
     \begin{subfigure}{0.32\textwidth}
         \centering
         \begin{overpic}[width=0.95\textwidth,trim=40 0 0 0,clip]{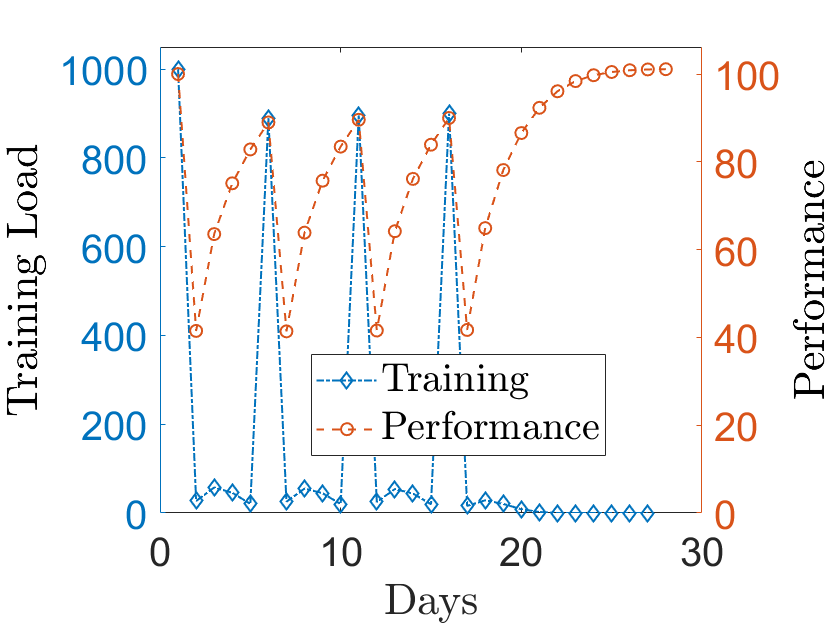}
         \put(-2,42){\rotatebox{90}{\makebox(0,0){\small Training impulse}}}
        \end{overpic}
         \caption{ }
         \label{subfig:g59_2}
     \end{subfigure}
     \begin{subfigure}{0.32\textwidth}
         \centering
         \begin{overpic}[width=0.95\textwidth,trim=40 0 0 0,clip]{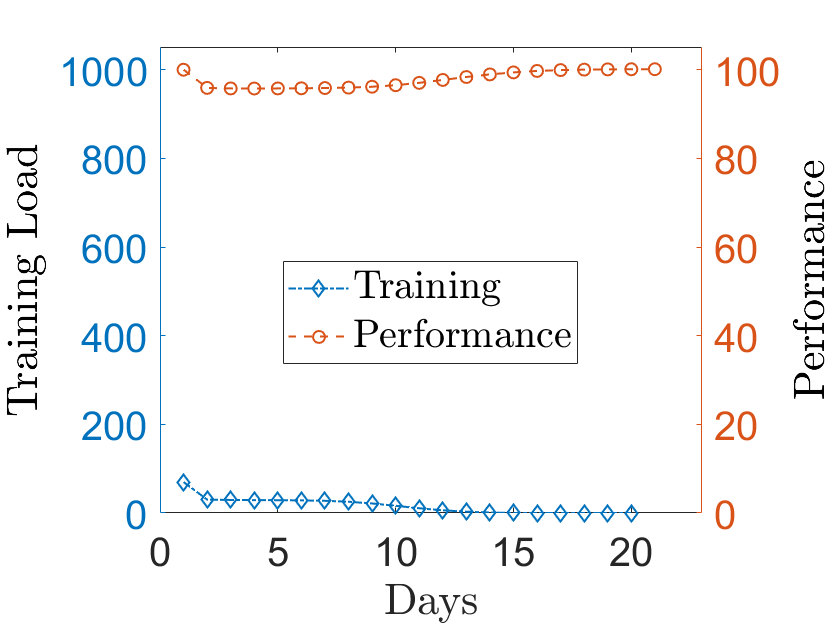}
         \put(-2,42){\rotatebox{90}{\makebox(0,0){\small Training impulse}}}
        \end{overpic}
         \caption{ }
         \label{subfig:g59_3}
     \end{subfigure}
        \caption{Optimized training impulse profiles to maximize performance on day 28 using the modified Fitness--Fatigue model with both power law diminishing returns and fatigue feedback \eqref{eq:doubleampdecay} for the parameters:
        (a) $k_3 = 0.25, k_4 =100$, (b) $k_3 = 0.25, k_4 =50$, 
        (c) $k_3 = 0.5, k_4 =10$ and all other parameters were:
    $P_0  = 100,
    k_1   = 0.01,
    k_2   = 0.1,
    \tau_1 = 10,
    \tau_2 = 2$.
    }
        \label{fig:periodizationAndTaperingSolutionswithDoubleAmp}
\end{figure}

This now allows us to put together a performance model that naturally displays both periodization and tapering in the optimized training impulse. Consider incorporating diminishing returns modeled by a power law and fatigue feedback, as in \eqref{eq:doubleampdecay}, for the same conditions as Fig.~\ref{subfig:g55_1}-\ref{subfig:g55_2} and we obtain those visible in Fig.~\ref{subfig:g59_1}-\ref{subfig:g59_2}. In Fig.~\ref{subfig:g59_1}, we can observe a training impulse profile that displays periodization, tapering, and some moderated training impulses that are not constrained by the lower or upper bound. Comparing Fig.~\ref{subfig:g55_1} to Fig.~\ref{subfig:g59_1}, we see a reduction in the variance of the training impulse profile and the addition of a taper. Note, in this context of modified magnitudes, if we increase the value of the parameters responsible for tapering and periodization too much, rather than obtaining a heavily periodized training program that displays tapering in the peaks, we more typically suppress the overall training effort, see Fig.~\ref{subfig:g59_3}. Fortunately, magnitude modifications are not the only means of mathematical intervention that we can have on the performance model; there are time-constant changes that we can make, which have the capacity to incentivize high training impulses.

In summary, under optimization, we saw that diminishing returns captured by an exponential decay, as given in \eqref{eq:MMexponential}, gave rise to the linear taper in \eqref{eq:linear-taper}, a power decay \eqref{eq:MMpower} gave rise to an exponential taper \eqref{eq:exponential-taper} and a logistic decay \eqref{eq:MMlogistic} gave rise to a step-down taper \eqref{eq:step-taper}. We also saw that if we assume that sustained fatigue has a deleterious effect on the current day fitness gain \eqref{eq:fatiguefeedback1} and \eqref{eq:doubleampdecay}, this naturally gave rise to periodization, Fig.~\ref{fig:periodizationSolutions}--\ref{fig:periodizationAndTaperingSolutionswithDoubleAmp}. When the two effects were combined, however, there is the possible issue that these suppress the model's capacity to derive benefit from large training impulses, and the optimum result is little to no training with a final performance similar to the initial performance, Fig.~\ref{subfig:g59_3}. We will now detail how time-constant changes can also give rise to varying behaviors in the optimal training profile solution and can incentivize relatively large training impulses.

\subsection{Time-Constant Changes} \label{Sec:Time-Constant Based Alterations}

Consider the idea that a small training impulse, such as a training impulse slightly more than a warm-up, will have a small and immediate positive impact on performance which lasts a relatively small amount of time. A large training impulse, on the other hand, will have a more significant effect which will initially be dominated by fatigue, take a few days for the response to reach a positive return and, overall, will persist for many times that of a small training impulse. 

This is partially captured in the modified magnitude case by diminishing returns. That is, the return on training could initially be dominated by a fitness gain, but as the training impulse increases and the return on additional training diminishes, fatigue may be the dominant immediate effect on the change in performance. 

Another avenue for emphasizing this is to make changes to the form of the time constant in the Fitness--Fatigue model. It is worth noting that the magnitude modifications and the time-constant changes are not different means to the same end. The magnitude response to training and how this decays in time are independent, orthogonal mechanisms which can each be isolated and observed independently (although, we do note that this is obscured and thus complicated by the daily training impulse assumption). 

In a similar fashion to the modified magnitude case, we will consider three possible forms of time-constant changes:
\begin{enumerate}
     \item Power increase in time-decay with training impulse 
     \begin{align} \label{eq:powerTCC}
         \tau_j(w(i)) = \tau_{j1} + \tau_{j2} w(i)^{m},
     \end{align}
     \item Logarithmic increase in time-decay with training impulse
     \begin{align} \label{eq:logarithmicTCC}
          \tau_j(w(i)) = \tau_{j1} + \tau_{j2} \ln(1+w(i)),
     \end{align}
     \item Logistic increase in time-decay with training impulse
     \begin{align} \label{eq:logisticTCC}
         \tau_j(w(i)) = \tau_{j1} + \frac{\tau_{j2}}{ 1 + \exp (-\tau_{j3} (w(i) - \tau_{j4}))},
     \end{align}
\end{enumerate}
where $\tau_{j1} \geq 0$ is the baseline time decay constant, $\tau_{j2} \geq 0$ controls how much the baseline time decay constant is elongated due to a training impulse, $\tau_{j3} > 0$ and $\tau_{j4} \geq 0$ are relevant to the logistic case and control the slope and offset of the function, respectively, and these functions appear in the performance equation as
\begin{align}
          P(n) = P_0 + \sum_{i=1}^{n-1} k_1 w(i) \exp \left( \frac{-(n-i)}{\tau_{1}(w(i)) } \right) - k_2 w(i) \exp \left( \frac{-(n-i)}{\tau_{2}(w(i)) } \right).
\end{align}
The form of the time-constant changes makes both the individual impulse responses and the overall performance equation transcendental functions, which do not lend themselves to analytical optimization. That is, even when the days can be considered independent and optimization of the series of training impulses can be reduced to the univariate case, the derivative of the impulse response with respect to $w$ yields an equation that, when set equal to zero, cannot be used to isolate $w$. This leaves us with two options: first, as we have been doing, using numerical optimization, or second, where possible, coming up with analytical bounds which enclose the optimal solution. 

\begin{figure}[ht!]
     \centering
     \begin{subfigure}{0.32\textwidth}
         \centering
         \begin{overpic}[width=0.95\textwidth,trim=40 0 0 0,clip]{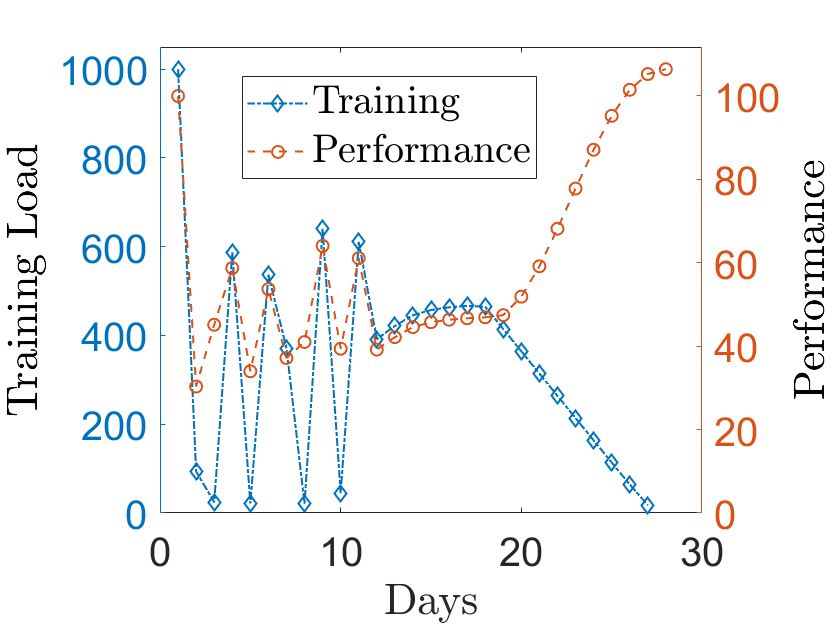}
                  \put(-2,42){\rotatebox{90}{\makebox(0,0){\small Training impulse}}}
        \end{overpic}
         \caption{ }
         \label{subfig:g56_1}
     \end{subfigure}
     \begin{subfigure}{0.32\textwidth}
         \centering
         \begin{overpic}[width=0.95\textwidth,trim=40 0 0 0,clip]{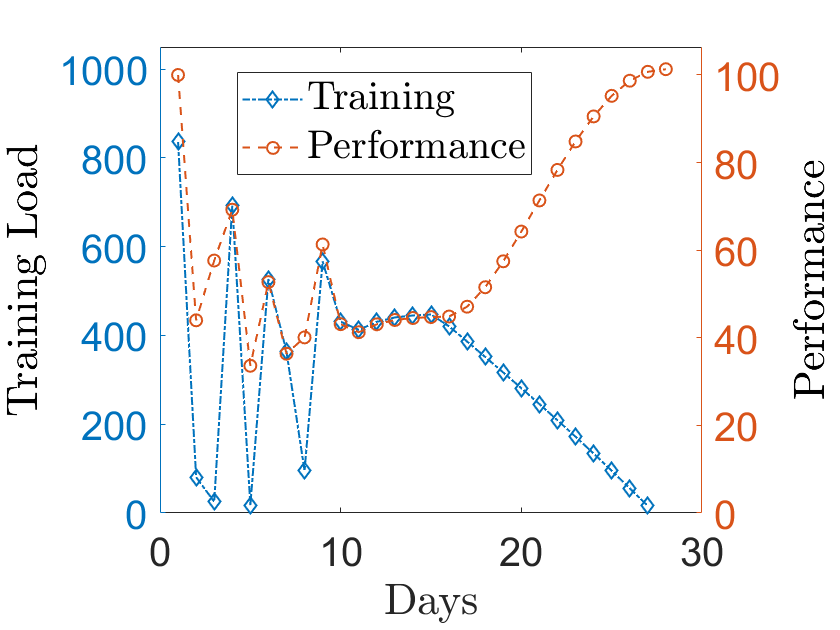}
                  \put(-2,42){\rotatebox{90}{\makebox(0,0){\small Training impulse}}}
        \end{overpic}
         \caption{ }
         \label{subfig:g56_2}
     \end{subfigure}
     \begin{subfigure}{0.32\textwidth}
         \centering
         \begin{overpic}[width=0.95\textwidth,trim=40 0 0 0,clip]{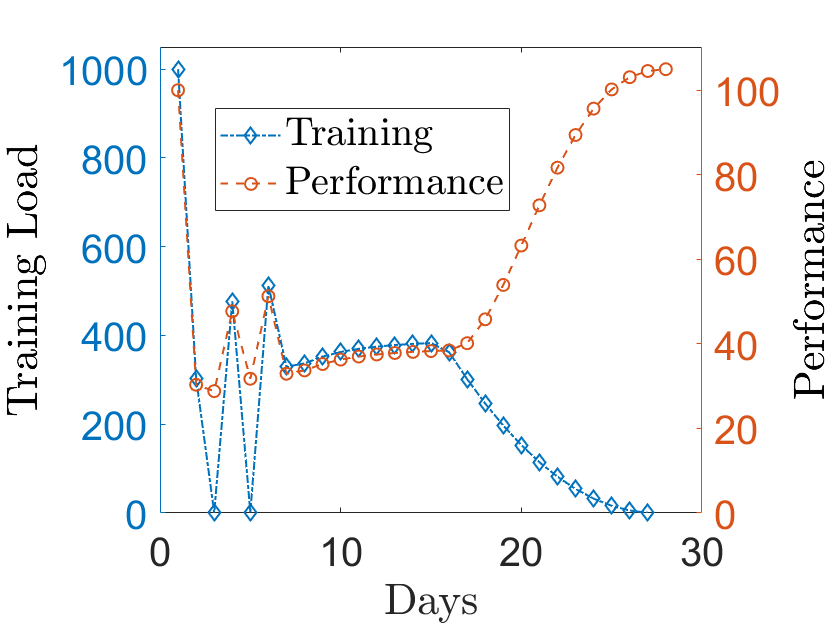}
                  \put(-2,42){\rotatebox{90}{\makebox(0,0){\small Training impulse}}}
        \end{overpic}
         \caption{ }
         \label{subfig:g56_3}
     \end{subfigure}
          \begin{subfigure}{0.32\textwidth}
         \centering
         \begin{overpic}[width=0.95\textwidth,trim=40 0 0 0,clip]{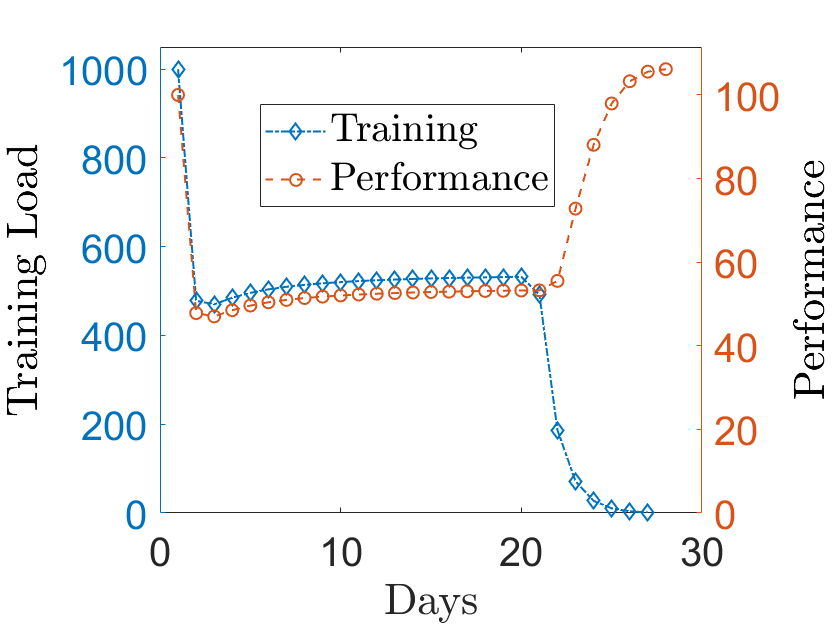}
                  \put(-2,42){\rotatebox{90}{\makebox(0,0){\small Training impulse}}}
        \end{overpic}
         \caption{ }
         \label{subfig:g56_4}
     \end{subfigure}
     \begin{subfigure}{0.32\textwidth}
         \centering
         \begin{overpic}[width=0.95\textwidth,trim=40 0 0 0,clip]{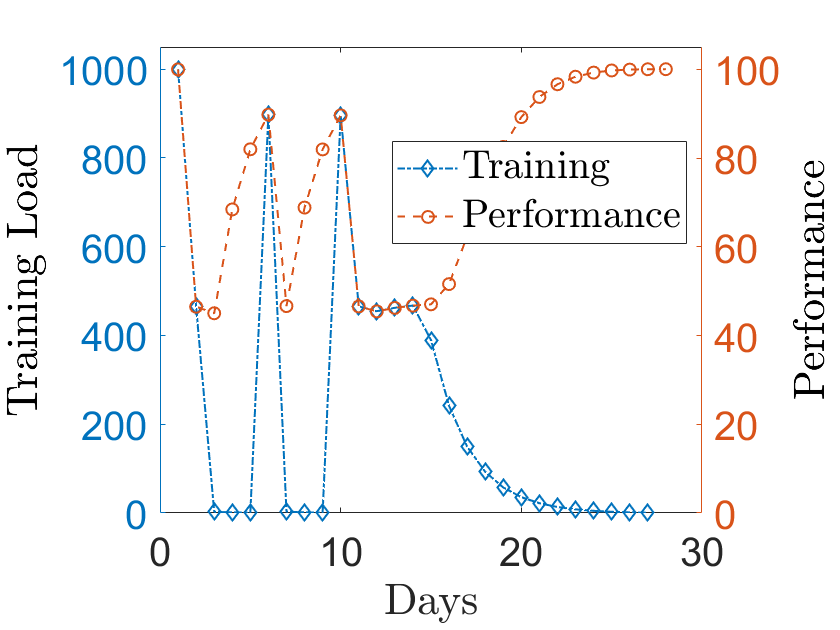}
                  \put(-2,42){\rotatebox{90}{\makebox(0,0){\small Training impulse}}}
        \end{overpic}
         \caption{ }
         \label{subfig:g56_5}
     \end{subfigure}
     \begin{subfigure}{0.32\textwidth}
         \centering
         \begin{overpic}[width=0.95\textwidth,trim=40 0 0 0,clip]{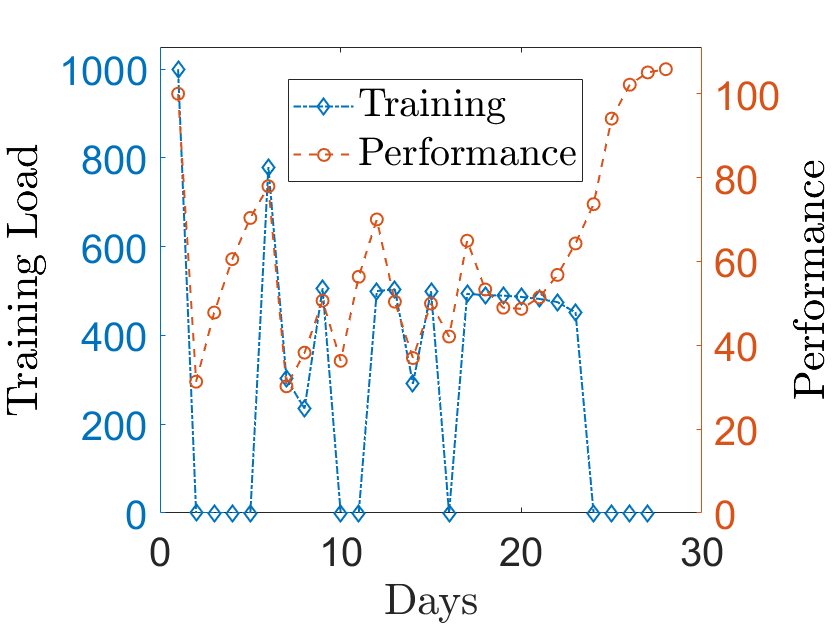}
                  \put(-2,42){\rotatebox{90}{\makebox(0,0){\small Training impulse}}}
        \end{overpic}
         \caption{ }
         \label{subfig:g56_6}
     \end{subfigure}
        \caption{Optimized training impulse profiles to maximize performance on day 28 using the modified Fitness--Fatigue model with the nonlinear time-constant changes and parameters:
        (a) power \eqref{eq:powerTCC}, $m=1, \tau_{11} = 1, \tau_{12} = 10/(P_0/k_2), \tau_{21} = 0.2, \tau_{22} = 4/(P_0/k_2)$,
        (b) same as (a) but with $\tau_{12} = 6/(P_0/k_2)$,
        (c) same as (a) but with $m=1/2$,
        (d) Logarithmic \eqref{eq:logarithmicTCC}, $\tau_{11} = 0, \tau_{12} = 6/\ln(P_0/k_2), \tau_{21} = 0, \tau_{22} = 2/\ln{(P_0/k_2)}$,
        (e) same as (d) but with $\tau_{12} = 3/\ln(P_0/k_2)$,
        (f) Logistic \eqref{eq:logisticTCC}, $\tau_{11} = 2, \tau_{12} = 10, \tau_{13} = \tau_{23} = 1/10, \tau_{14} = \tau_{24} = 500, \tau_{21} = 1, \tau_{22} = 3$,
        and all other parameters were:
    $P_0  = 100,
    k_1   = 0.01,
    k_2   = 0.1$. 
    }
        \label{fig:TimeConstnatChanges}
\end{figure}

For the first option, observe a selection of numerically obtained optimal training impulse profiles which maximize the performance after 28 days in Fig.~\ref{fig:TimeConstnatChanges}. Here, we can see that, by having larger training impulses decay more slowly on average, we can motivate periodization in the optimized training impulse which then stabilizes and tapers into the event in a linear way Fig.~\ref{subfig:g56_1}-\ref{subfig:g56_2}, parabolic way Fig.~\ref{subfig:g56_3}, exponential way Fig.~\ref{subfig:g56_4}-\ref{subfig:g56_5} or in a step-down way Fig.~\ref{subfig:g56_6}. Moreover, when comparing Fig.~\ref{subfig:g56_1}-\ref{subfig:g56_2}, we can see an increased taper (beginning day 19 compared to day 16) being driven by a reduction in $\tau_{12}$. We also see this being the case in the comparison of Fig.~\ref{subfig:g56_4}-\ref{subfig:g56_5}. Particularly noticeable in the comparison of Fig.~\ref{subfig:g56_4}-\ref{subfig:g56_5} is that we see an increase in the periodization of the training impulse for a reduction in $\tau_{12}$. This is because, if $O(\tau_1(w_{\text{max}})) \gg n$, then medium training impulses persist to the end date about as well as large training impulses. Whereas, if $O(\tau_1(w_{\text{max}})) \lesssim n$, then we require large training impulses during the initial stages of training to have a relevant impact on day $n$.

Analytically speaking, we can get a sense of the optimal tapering dynamics by calculating an upper bound on the training impulse. We can do this by finding the training impulse which has its $t_r$ as the end date (any more training than this would yield a negative impact on the final performance). To calculate the time to return, we can set the positive and negative impulse responses as equal to each other and, in this instance, set the time to return equal to the time to the maximum performance date $j$ and solve for the resulting training impulse upper bound $w_u(j)$
\begin{align}
    k_1 w_u(j) \exp \left( \frac{-j}{\tau_{1}(w_u(j)) } \right) &= k_2 w_u(j) \exp \left( \frac{-j}{\tau_{2}(w_u(j)) } \right) \nonumber \\
    \left(\frac{1}{\tau_{2}(w_u(j))} - \frac{1}{\tau_{1}(w_u(j)) } \right) &= \frac{1}{j} \ln \left( \frac{k_2}{k_1} \right)
\end{align}
where we have assumed $w_u(j) \neq 0$. From here, we will need to consider each form of time-constant change: power, logarithmic and logistic, as defined in \eqref{eq:powerTCC}-\eqref{eq:logisticTCC}, respectively. For the power case, when $\tau_j(w_u(j)) \approx \tau_{j2} w_u(j)^{m}$, this yields the approximate upper limit on the training impulse as we taper into an event as
\begin{align}
   w_u(j) \approx \left( j \left(\frac{\tau_{12} - \tau_{22}}{\tau_{12}\tau_{22}}\right) \frac{1}{\ln( k_2/k_1)} \right)^{1/m}.
\end{align}
This suggests that as we approach the date of our desired maximum performance, we should decrease our training impulse in an approximately $j^{1/m}$ fashion, where, again, the above is not the taper itself but an upper bound beyond which training is definitely not beneficial. This is consistent with what was observed in the numerical optimization, Fig.~\ref{subfig:g56_1}-\ref{subfig:g56_3}, which has a linear taper, another linear taper and a parabolic taper for the cases $m=\{1,1,1/2\}$, respectively. For the logarithmic case, when $\tau_j(w_u(j)) \approx \tau_{j2} \ln(1 + w_u(j))$, this yields the approximate upper limit
\begin{align}
    w_u(j) \approx \exp \left( j \left(\frac{\tau_{12} - \tau_{22}}{\tau_{12}\tau_{22}}\right) \frac{1}{\ln( k_2/k_1)} \right) - 1.
\end{align}
This result suggests that we would want to decrease our training impulse in an exponential fashion as we approach the date of our desired maximum performance. Comparing this to the numerical results, we can see that Fig.~\ref{subfig:g56_4}-\ref{subfig:g56_5} also display a decaying exponential taper. The logistic case is more complicated than the previous two and it is difficult to make general statements. Some tentative observations include, if $\tau_{j3} \ll 1$, then the logistic curve can behave similarly to a linear power and the optimal training impulse solution can display a linear taper; if $O(\tau_{j3}) = 1$ or $\tau_{j3} \gg 1$, then we can observe a sharp change in time-constant at the threshold training value $\tau_{j4}$ and training might be prioritized as above this threshold---although, the optimized distribution of training impulse will still vary widely for each of these regimes depending on the values of $\tau_{j1}$ and $\tau_{j2}$, or even $k_1$ and $k_2$.

In summary, time-constant changes offer us another perspective on why we may choose to periodize and taper a training program for an upcoming event. In this case, relatively large training impulses are incentivized by being rewarded with fitness gains which persist for longer periods of time---which is fundamentally different to the mechanism that drove periodization in the modified magnitude case where sustained fatigue offset some of the fitness gain of consecutive training impulses. In the case of the changed time-constant, tapering is facilitated by training impulse responses having a decreasing time to return with decreasing training impulses.

\subsection{Combined Changes} \label{Sec:Combined Changes}

Up to this point, we have considered three modifications to the Fitness--Fatigue model:
\begin{enumerate}
    \item Diminishing returns on increasing training impulses. This allows for a small training impulse to have an immediate performance gain, while a large training impulse may have an immediate performance loss.
    \item Fatigue sustained on previous days reduces not just the magnitude of the allowable training impulse through the upper bound $k_2 w(i) \le P(i)$, but also the amount of fitness gain for the same training impulse. 
    \item Large training impulses produce changes in both the fitness and fatigue response that decay more slowly than a small training impulse.
\end{enumerate}

\begin{figure}[ht!]
     \centering
     \begin{subfigure}{0.32\textwidth}
         \centering
         \begin{overpic}[width=0.95\textwidth,trim=40 0 0 0,clip]{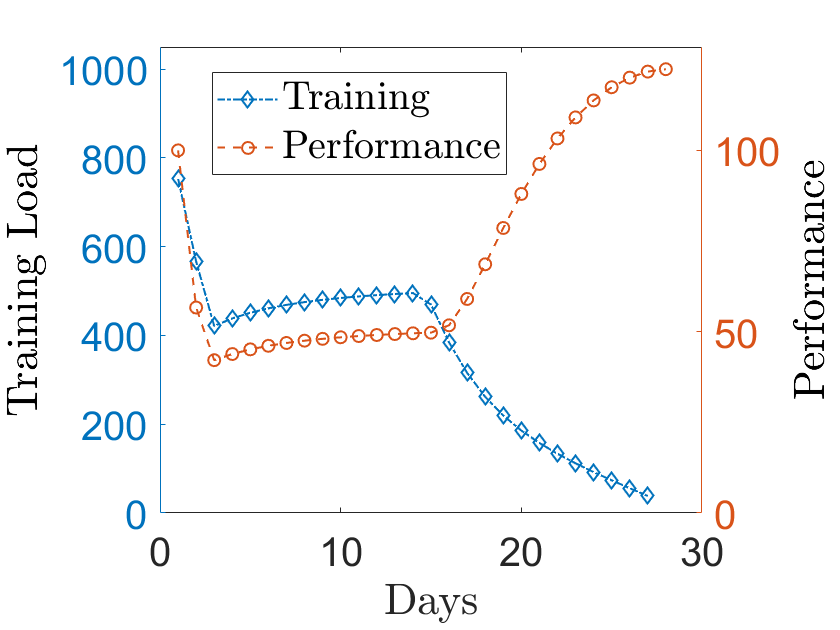}
         \put(-2,42){\rotatebox{90}{\makebox(0,0){\small Training impulse}}}
        \end{overpic}
         \caption{ }
         \label{subfig:g60_1}
     \end{subfigure}
     \begin{subfigure}{0.32\textwidth}
         \centering
         \begin{overpic}[width=0.95\textwidth,trim=40 0 0 0,clip]{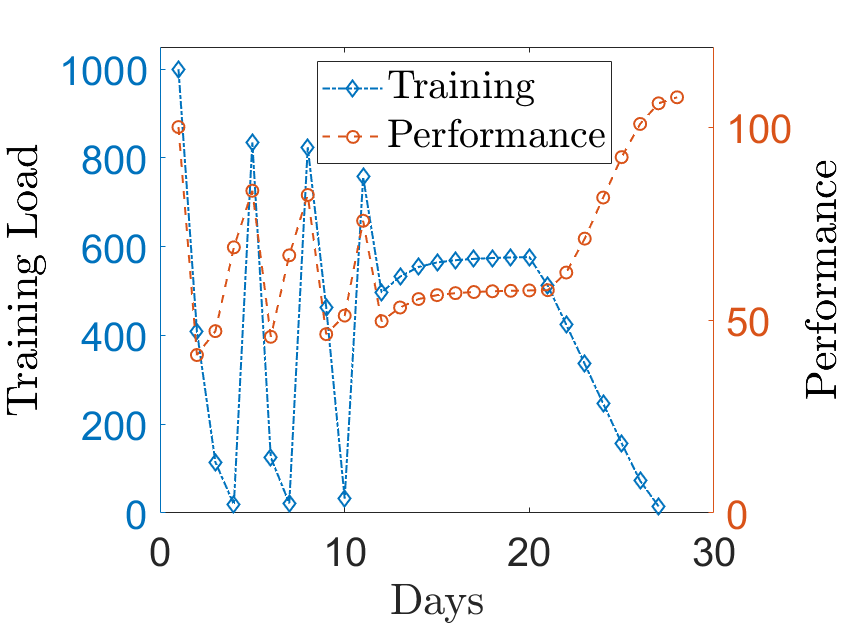}
         \put(-2,42){\rotatebox{90}{\makebox(0,0){\small Training impulse}}}
        \end{overpic}
         \caption{ }
         \label{subfig:g60_2}
     \end{subfigure}
     \begin{subfigure}{0.32\textwidth}
         \centering
         \begin{overpic}[width=0.95\textwidth,trim=40 0 0 0,clip]{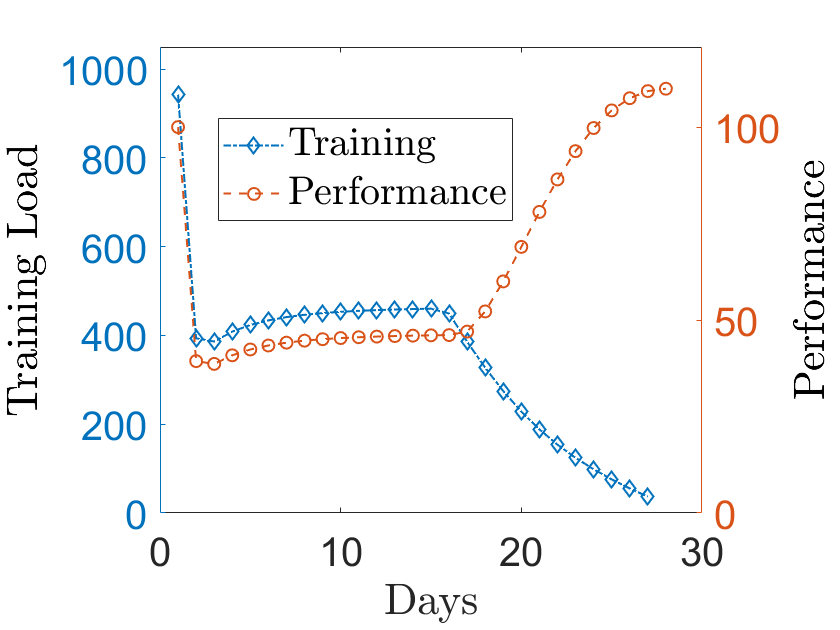}
         \put(-2,42){\rotatebox{90}{\makebox(0,0){\small Training impulse}}}
        \end{overpic}
         \caption{ }
         \label{subfig:g60_3}
     \end{subfigure}
          \begin{subfigure}{0.32\textwidth}
         \centering
         \begin{overpic}[width=0.95\textwidth,trim=40 0 0 0,clip]{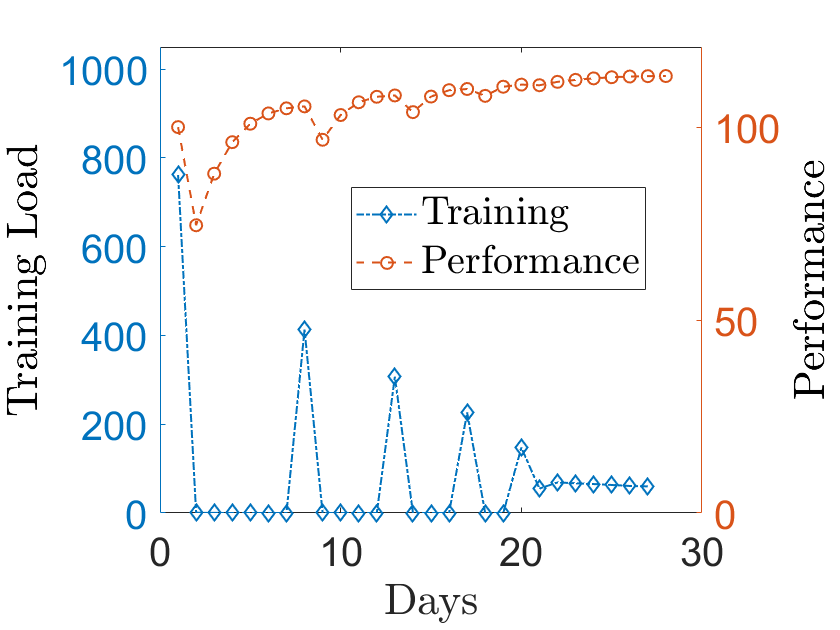}
         \put(-2,42){\rotatebox{90}{\makebox(0,0){\small Training impulse}}}
        \end{overpic}
         \caption{ }
         \label{subfig:g60_4}
     \end{subfigure}
     \begin{subfigure}{0.32\textwidth}
         \centering
         \begin{overpic}[width=0.95\textwidth,trim=40 0 0 0,clip]{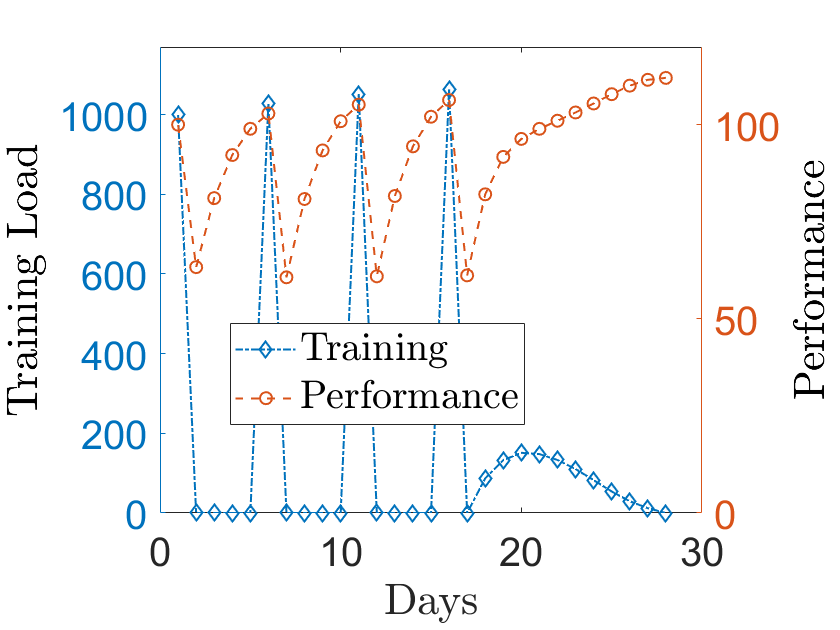}
         \put(-2,42){\rotatebox{90}{\makebox(0,0){\small Training impulse}}}
        \end{overpic}
         \caption{ }
         \label{subfig:g60_5}
     \end{subfigure}
     \begin{subfigure}{0.32\textwidth}
         \centering
         \begin{overpic}[width=0.95\textwidth,trim=40 0 0 0,clip]{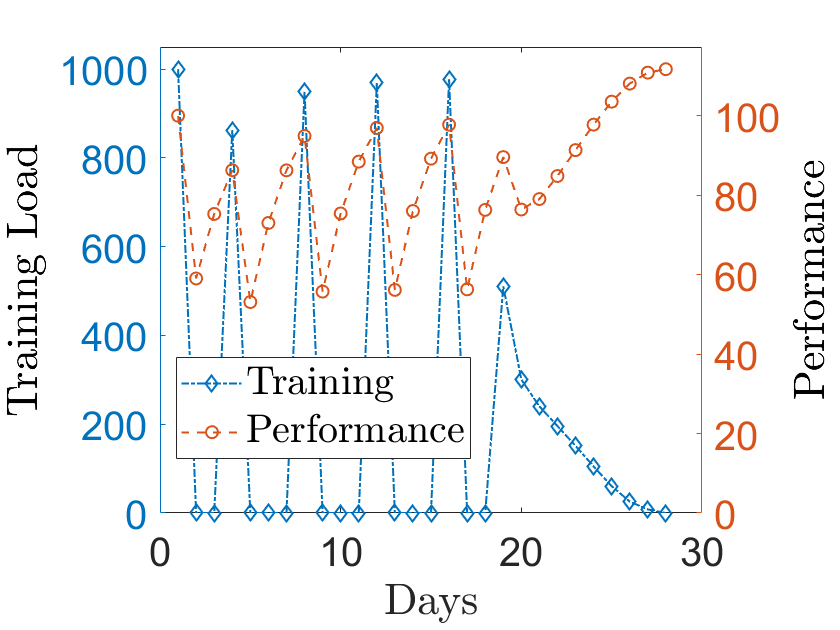}
         \put(-2,42){\rotatebox{90}{\makebox(0,0){\small Training impulse}}}
        \end{overpic}
         \caption{ }
         \label{subfig:g60_6}
     \end{subfigure}
        \caption{Optimized training impulse profiles to maximize performance on day 28 using \eqref{eq:CombinedChanges} with nonlinear magnitude modifications, time-constant changes and parameters:
        (a) Exponential decay \eqref{eq:MMexponential}, logarithmic increase \eqref{eq:logarithmicTCC} $k_3 = 50, \tau_{11} = 1, \tau_{12} = 6/\ln(P_0/k_2), \tau_{21} = 0.5, \tau_{22} = 2/\ln(P_0/k_2)$,
        (b) Power decay \eqref{eq:MMpower}, linear increase \eqref{eq:powerTCC}, $k_3 = 0.5, \tau_{11} = 1, \tau_{12} = 6/(P_0/k_2), \tau_{21} = 0.5, \tau_{22} = 2/(P_0/k_2)$,
        (c) Exponential decay \eqref{eq:MMexponential}, square root increase \eqref{eq:powerTCC}, $k_3 = 50, \tau_{11} = 1, \tau_{12} = 6/\sqrt{P_0/k_2}, \tau_{21} = 1, \tau_{22} = 2/\sqrt{P_0/k_2}$,
        (d) Exponential decay \eqref{eq:MMexponential}, logarithmic increase \eqref{eq:logarithmicTCC}, $k_3 = 200, k_4 = 20, \tau_{11} = 0.5, \tau_{12} = 6/\ln(P_0/k_2), \tau_{21} = 0.5, \tau_{22} = 2/\ln(P_0/k_2)$,
        (e) Power decay, fatigue-feedback \eqref{eq:doubleampdecay} and logarithmic increase \eqref{eq:logarithmicTCC}, $k_3 = 0.25, k_4 = 60, \tau_{11} = 0.5, \tau_{12} = 6/\ln(P_0/k_2), \tau_{21} = 0.5, \tau_{22} = 2/\ln(P_0/k_2)$,
        (f) Power decay, fatigue-feedback \eqref{eq:doubleampdecay}, square root increase \eqref{eq:powerTCC}, $k_3 = 0.25, k_4 = 125, \tau_{11} = 1, \tau_{12} = 6/\sqrt{P_0/k_2}, \tau_{21} = 1, \tau_{22} = 2/\sqrt{P_0/k_2}$,
        and all other parameters are: $P_0 = 100, k_1 = 0.15, k_2 = 0.1$.
     }
      \label{fig:CombinedChanges}  
\end{figure}

Having examined the three modifications separately, we now consider representative combinations of diminishing returns, fatigue feedback, and training impulse dependent time constants. Fig.~\ref{fig:CombinedChanges} shows several optimal training impulse profiles and illustrates that it is possible to elicit: relatively consistent training and tapering profiles, Fig.~\ref{subfig:g60_1} and~\ref{subfig:g60_3}; periodized profiles, Fig.~\ref{subfig:g60_4}-\ref{subfig:g60_6}; as well as a combination of periodized and relatively consistent training and with a taper, Fig.~\ref{subfig:g60_2}.

\begin{figure}[ht!]
 \centering
         \begin{overpic}[width=0.75\textwidth,trim=80 0 0 0,clip]{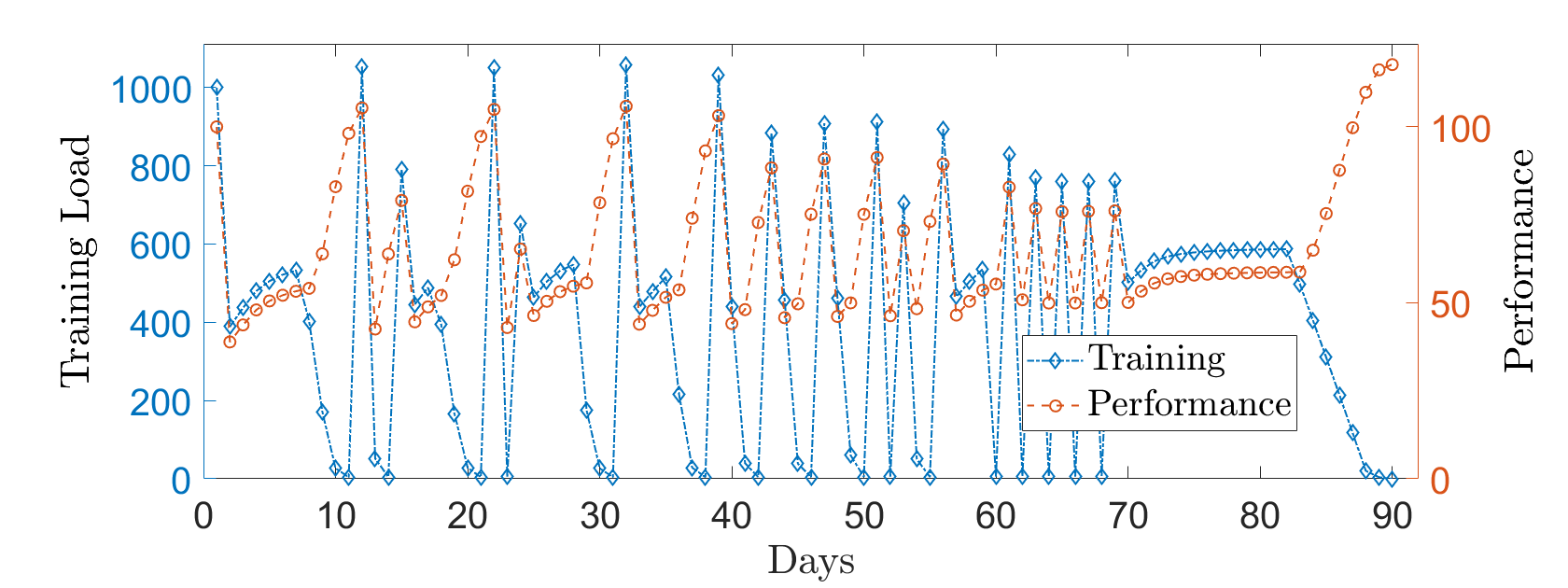}
         \put(-1,22){\rotatebox{90}{\makebox(0,0){Training impulse}}}
        \end{overpic}
        \caption{Optimized training impulse profile to maximize performance on day 90 using the modified Fitness--Fatigue Impulse Response model with a power law decay in the fitness magnitude, linear increase in the decay time-constant and parameters: $k_1 = 0.15, k_2 = 0.1, k_3 = 0.5, \tau_{11} = 1, \tau_{12} = 1/100, \tau_{21} = 0.75, \tau_{22} = 1/500$.}
    \label{Fig:90 Days}
\end{figure}

We note that the optimized training impulse profiles considered thus far have been limited to a 28-day time frame, whereas practical training programs often span considerably longer periods. For an example of an optimized training impulse profile over 90 days, see Fig.~\ref{Fig:90 Days}. This is for the case that we have a power law decay in the fitness magnitude, and have a linear increase in the impulse response decay-constant with respect to the training impulse.

Examining Fig.~\ref{Fig:90 Days}, the resulting profile displays several phases with differing oscillatory structure (approximately 1–40, 40–60, and 60–70 days), followed by a period of relatively stable training and then a gradual taper. While these `block' phases bear a qualitative resemblance to patterns commonly associated with training periodization, this similarity should be interpreted cautiously: as optimized solutions over long time frames which display periodization are particularly susceptible to the presence of multiple local optima.

In summary, we were able to obtain a variety of complex optimized training programs under different functional forms and parameter selections in the combined model \eqref{eq:CombinedChanges}. In some cases, the optimal solution displayed relatively consistent training followed by a taper, while in others it exhibited periodization with varying magnitudes and frequencies. Moreover, with an established relationship between magnitude and time-constant changes that give rise to different tapers, this opens up a new avenue for answering the question of optimal taper. If we can establish what type of diminishing returns or time-constant elongation applies in practice, we will also contribute to answering which tapering pattern is optimal. We hope that this mathematical variety has the capacity to reflect the realistic diversity observed in practice across different sports and individual athletes.

\section{Discussion} \label{sec:Discussion}

\subsection*{Statistical Inference of Parameters}

For optimization, and indeed for most predictive applications, Fitness–Fatigue Impulse Response models are useful only if two conditions are satisfied. First, the model parameters must be inferable from data. Second, the model must capture the relevant physiological effects with sufficient fidelity that its predictions remain consistent with established empirical understanding.

Statistical deficiencies in parameter estimation have been documented for the Banister model and closely related Fitness–Fatigue Impulse Response formulations \citep{hellard2006assessing,busso2006using,marchal2025statistical}. Our analysis further shows that the Banister model fails to satisfy the second condition, as it does not yield optimal recommendations consistent with empirically derived training guidelines. These results indicate that the Banister model, and by extension the current implementations of Fitness–Fatigue Impulse Response-type models, do not yet fulfill either criterion.

The present work does not directly address the first condition. Nevertheless, it offers two indirect contributions relevant to parameter estimation. First, we derived the condition in \eqref{eq:BanisterParameterConstrain} that separates bounded and unbounded regimes of the Banister model which can assist in restricting the feasible parameter space during estimation. Second, the broader message of our work---that realistic optimal behavior is a necessary property of a practical model---provides a conceptual framework for evaluating and improving future model formulations. For example, while a fitness-only model may mitigate some of the statistical inference difficulties reported in the literature, such a model would lack the structural capacity required to support optimization-based analyses.

Importantly, the arguments presented here are not undermined by the statistical shortcomings identified in previous parameter estimation studies. Regardless of whether the parameters of standard Fitness--Fatigue Impulse Response models can be reliably inferred, our analysis shows that these models would still fail to produce optimal training impulse profiles consistent with empirically derived guidelines. Note, this is not to suggest that Fitness--Fatigue Impulse Response models will never satisfy the two conditions for practical optimization use; rather, it remains the subject of future work. One promising direction for addressing the parameter estimation challenges is the use of Bayesian inference \citep{peng2024bayesian,marchal2025statistical}, which allows prior beliefs to be incorporated. Recent work by \citet{marchal2025statistical} indicates, however, that such approaches may require careful implementation, as posterior distributions for certain parameters may exhibit limited updating from their priors. 

Finally, we make a comment regarding the accurate estimation of the parameters introduced in this paper, such as those governing diminishing returns or time-decay elongation. These may require the development of dedicated empirical protocols capable of isolating and sampling these effects. It is unlikely that these parameters could be reliably inferred from periodic performance assessments conducted on, for example, weekly or fortnightly timescales.


\subsection*{Mechanistic vs. Data-driven Modeling Approaches}

The present work is concerned with mechanistically motivated models of training and performance. The principal reason is that such models impose an explicit mathematical structure on the relationship between training inputs and performance outcomes, with assumptions and parameters that remain interpretable. This makes them a natural setting for analytical optimization. When an optimized training profile appears plausible, or implausible, the source of that behavior can be traced to identifiable aspects of the model.

This focus should be viewed against the broader development of the field. Recent work has increasingly explored data-driven alternatives, including regression-based approaches \citep{avalos2003modeling}, latent-variable models \citep{weaving2021latent}, and machine-learning methods \citep{imbach2022training,imbach2022use}. These approaches are an important part of the current modeling landscape, particularly when sufficiently rich datasets are available. At the same time, data-driven and mechanistic approaches need not be regarded as competing paradigms. They address related questions through different representational frameworks and may ultimately prove complementary.

Within that broader picture, the present paper has a more specific aim. It examines the consequences of optimizing established mechanistic models and identifies which modeling features appear necessary if the resulting training profiles are to better reflect realistic tapering or periodization.

\subsection*{Realism Issues in Fitness–Fatigue Impulse Response Models}

The primary aim of this study was to demonstrate that standard Fitness–Fatigue Impulse Response models lack the incorporation of sufficient physiological effects to produce optimization outcomes consistent with empirically derived training guidelines. Although several modifications were proposed to illustrate how greater realism might be achieved, the present analysis does not attempt to reproduce the real training profile of any particular athlete.

A fully realistic reproduction of athlete training profiles was not pursued for several reasons. First, while empirical studies consistently support the inclusion of periodization and tapering as components of effective training programs \citep[see, for example,][]{turner2011science}, the literature provides insufficient evidence to specify what constitutes an optimal implementation of these principles \citep[see][]{vachon2021effects}. Second, constructing a model tailored to a specific athlete would unduly narrow the scope of the study and obscure its general theoretical insights. Third, achieving full realism would require incorporating additional physiological effects---such as, for example, injury risk---whose accurate characterization remains an active and complex field of research, warranting its own dedicated treatment.

Our modifications to the Fitness--Fatigue formulation addressed some realism issues, such as the absence of periodization, the lack of nontrivial tapering, and unrealistic magnitudes of performance change. Others remain unresolved, including unrealistically large training impulses, especially those on day one, and substantial decrements in performance resulting from excessive fatigue accumulation. 

In summary, the present work should be regarded as a step toward reconciling theoretical performance models with empirically grounded training principles. While the proposed extensions mitigate some realism issues, further theoretical refinement and empirical validation are needed before Fitness–Fatigue Impulse Response models can yield fully realistic and practically optimizable representations of athletic performance.

\section{Conclusion} \label{sec:Conclusion}

To enhance our understanding of athletic performance, it is essential that trends in the optimized training impulse derived from our mathematical models match observations documented in the experimental sports science literature. Having an accurate model of athletic performance will enable us to determine the training impulse profile that maximizes performance improvement by a specified date, or, conversely, the minimal training impulse profile required to reach a desired level of improvement within a given timeframe. In this way, a robust quantitative model could inform both elite athletes seeking peak performance and time-constrained individuals aiming for efficient progress. In addition, an accurate quantitative tool will help people set realistic expectations regarding the time frames and amount of performance improvement an investment in training will achieve. Furthermore, the process of developing and interpreting such a model may itself yield valuable conceptual understanding and intuition for coaches, sport scientists, and practitioners. 

To induce a linear Fitness--Fatigue Impulse Response model with exponential decay to display things such as nuanced tapering strategies and periodization we considered the nonlinear mathematical features:
\begin{enumerate}
    \item A fitness gain with diminishing returns with increasing training impulse. We found that if the diminishing returns were exponential, as expressed in \eqref{eq:MMexponential}, then the optimized training impulse arising from the impulse response model naturally adopted the linear taper defined in \eqref{eq:linear-taper}. If it were a power law decay, as shown in \eqref{eq:MMpower}, then it displayed the exponential taper defined in \eqref{eq:exponential-taper}. And if it were a logistic decay, as in \eqref{eq:MMlogistic}, it had the step-down taper defined in \eqref{eq:step-taper}.  
    
    \item Fatigue feedback reduces not just the quantity of allowable training but also the amount of fitness gain for the same training impulse, see equations \eqref{eq:fatiguefeedback1} and \eqref{eq:doubleampdecay}. This induced periodization to appear in the optimized training impulse. Although, this was not the only intervention to produce periodization in the optimized training impulse. 
    
    \item Large training impulses produce a performance change that has an impulse response with a slower decay than smaller training impulses. We found that if this change in the time-constant governing the impulse response decay was linear with respect to the training impulse, as expressed in \eqref{eq:powerTCC}, then it had the capacity to motivate periodized training and prompted a linear taper in the optimized training impulse. If the change in time constant were logarithmic, as given by \eqref{eq:logarithmicTCC}, it had reduced capacity to motivate periodization but still offered some incentive for high training impulses, and prompted an exponential taper. If it were logistic, as described in \eqref{eq:logisticTCC}, it incentivized training above a certain threshold whenever possible, and prompted a step-down taper. 
    
\end{enumerate}
We emphasize that the functions suggested for magnitude modifications and time-constant changes are merely for the point of demonstration and the choice of which one, mixture or entirely new function must be made based on whichever reflects the experimental evidence most appropriately. 

We believe that the combined magnitude decay and time-constant increase with increasing training impulse provides an illuminating perspective on how someone can reduce their fatigue in anticipation of an event whilst still performing some training. We also believe that fatigue-feedback and the desire for long-lasting performance improvements are two good reasons to periodize a training program.
That said, the concept of injury prevention and its impact on the formulation of real training programs warrants future investigation.


 
\appendix 
\renewcommand{\thesection}{Appendix \Alph{section}}

\section{Proof of Theorem \ref{theorem1}} \label{sec: Appendix}

To prove that $\bm{w}^*$ given in \eqref{eq:TrainAsHardASPossible} solves the optimization problem \eqref{eq:BanisterOptimizationProblem}, subject to the lower bound and upper bound constraints
\[
    c_i(\bm{w}) = -w(i) \le 0,
    \qquad
    h_i(\bm{w}) = k_2 w(i) - P(i) \le 0,
    \qquad i \in \{1,\dots,n-1\},
\]
we show that it satisfies the Karush--Kuhn--Tucker (KKT) conditions \citep{bertsekas2016nonlinear}. We denote the dual variables or KKT multipliers associated with the lower-bound constraints $c_i$ by $\nu_i$, and those associated with the upper-bound constraints $h_i$ by $\mu_i$.

\medskip
We first introduce the following reparameterization, which will be used throughout the remainder of the proof:
\begin{align}\label{eq:Reparameterization}
    A = k_1 e^{-t_r/\tau_1} = k_2 e^{-t_r/\tau_2}, \qquad
    &a = e^{-1/\tau_1}, \quad b = e^{-1/\tau_2}, \nonumber \\
    r = \frac{k_1}{k_2} \in (0,1), \qquad
    &z = \frac{b}{a} \in (0,1),
\end{align}
and define $\alpha = \exp\!\left(-\frac{\trp - t_r}{\tau_1}\right)$, $\beta  = \exp\!\left(-\frac{\trp - t_r}{\tau_2}\right),$ which satisfies $1 > \alpha > \beta > 0$. We also define the sequence of polynomials $Q_k(x)$ for $k \in \{1,2,\dots,n-\trp\}$ by the recurrence
\begin{align}\label{eq:BigBoiQ}
    Q_k(x) = x^{k-1} + \sum_{q=1}^{k-1} (r - z^q)\, Q_{k-q}(x),
    \qquad Q_1(x) = 1, \quad x \in [0,1].
\end{align}

\medskip
With this notation, the KKT multipliers are given explicitly by
\begin{align}
\nu_{i} = \label{eq:Nus}
\begin{cases}
 0, 
& 1 \le i \le n - t_r^+, \\[5pt]
\dfrac{-I(n - i)}{k_2},
& n - t_r^+ < i \le n-1,
\end{cases}
\end{align}
where $I(t) = k_1\exp(-t/\tau_1) - k_{2}\exp(-t/\tau_2)$ and
\begin{align}
\mu_{i} = \label{eq:Mus}
\begin{cases}
 \dfrac{A a^{i-1}}{k_2}\left( \alpha Q_i(1) - \beta Q_i(z) \right),
& 1 \le i \le n - t_r^+,  \\[5pt]
0, & n - t_r^+ < i \le n-1.
\end{cases}
\end{align}
Specifically, we show that $\bm{w}^*$ satisfies:
\begin{enumerate}[label=(\roman*)]

    \item \textbf{Primal Feasibility.}
    \begin{align}
        c_i(\bm{w}^*) \le 0
        \quad\text{and}\quad
        h_i(\bm{w}^*) \le 0,
        \qquad i \in \{ 1,\dots,n-1 \}.
    \end{align}

    \item \textbf{Complementary Slackness.}
    \begin{align}
        \nu_i\, c_i(\bm{w}^*) = 0
        \quad\text{and}\quad
        \mu_i\, h_i(\bm{w}^*) = 0,
        \qquad i \in \{ 1,\dots,n-1 \}.
    \end{align}

    \item \textbf{Stationarity.}
    For each $\ell \in \{1,\dots,n-1\}$,
    \begin{align} \label{eq:Stationarity}
        - \partial_{w(\ell)} P(\bm{w}^*,n)
        + \sum_{i =1}^{n-1}
            \Bigl[
                \nu_i\, \partial_{w(\ell)} c_i(\bm{w}^*)
              + \mu_i\, \partial_{w(\ell)} h_i(\bm{w}^*)
            \Bigr]
        = 0.
    \end{align}

    \item \textbf{Dual Feasibility.}
    \begin{align}
        \nu_i \ge 0
        \quad\text{and}\quad
        \mu_i \ge 0,
        \qquad i \in \{ 1,\dots,n-1 \}.
    \end{align}
\end{enumerate}

\subsection*{Proof of (i)}
The constraints are
\[
    c_i(\bm{w}) = -w(i) \le 0,
    \qquad
    h_i(\bm{w}) = k_2 w(i) - P(i) \le 0.
\]
For each day $i \in \{1,\dots,n-\trp\}$, the solution $\bm{w}^*$ activates the upper bound in \eqref{con:trainingconstraintUpper}, in which case $h_i(\bm{w}^*) = 0$ and $c_i(\bm{w}^*) < 0$. It then activates the lower bound in \eqref{con:trainingconstraintLower}, in which case $c_i(\bm{w}^*) = 0$ and $h_i(\bm{w}^*) < 0$, for $i \in \{n-\trp + 1 , \dots , n-1\}$. Hence
\[
    c_i(\bm{w}^*) \le 0
    \quad\text{and}\quad
    h_i(\bm{w}^*) \le 0,
    \qquad i \in \{ 1,\dots,n-1 \},
\]
so primal feasibility holds.

\subsection*{Proof of (ii)}
From $\bm{w^*}$ in \eqref{eq:TrainAsHardASPossible}, the upper bound is active for all
$i \in \{1,\dots,n - \trp \}$, so $ h_i(\bm{w}^*) = 0$,  $c_i(\bm{w}^*) < 0$. For these indices, the multipliers defined in \eqref{eq:Nus}  and \eqref{eq:Mus} satisfy $\nu_i = 0$ and $\mu_i > 0$, and thus
\[
    \nu_i\, c_i(\bm{w}^*) = 0,
    \qquad
    \mu_i\, h_i(\bm{w}^*) = 0.
\]
For the remaining days $i \in \{n - \trp + 1,\dots,n-1\}$, the lower bound is active, giving $c_i(\bm{w}^*) = 0,$ and $h_i(\bm{w}^*) < 0.$ Here \eqref{eq:Nus} gives $\nu_i > 0$ and \eqref{eq:Mus} yields $\mu_i = 0$, so
\[
    \nu_i\, c_i(\bm{w}^*) = 0,
    \qquad
    \mu_i\, h_i(\bm{w}^*) = 0.
\]
Thus, complementary slackness holds for every $i \in \{1,\dots,n-1\}$.

\subsection*{Proof of (iii)}
The equation for stationarity (iii) leads to the matrix equations
\begin{align} \label{eq:matrix 1 KKT}
\begin{bmatrix}
k_2 & -I(1) & -I(2) & \cdots & -I(n-t_r^+-1) \\
0   & k_2   & -I(1) & \cdots & -I(n-t_r^+-2) \\
0   & 0     & k_2   & \cdots & -I(n- t_r^+-3) \\
\vdots & \vdots & \vdots & \ddots & \vdots \\
0   & 0     & 0     & \cdots & k_2
\end{bmatrix}
\begin{bmatrix}
\mu_1 \\
\mu_2 \\
\mu_3 \\
\vdots \\
\mu_{(n - \trp)}
\end{bmatrix}
=
\begin{bmatrix}
I(n-1) \\
I(n-2) \\
I(n-3) \\
\vdots \\
I(t_r^+)
\end{bmatrix}
\end{align}
and
\begin{align} \label{eq:matrix 2 KKT}
\begin{bmatrix}
-k_2 & 0 &  \cdots & 0 \\
0   &- k_2    & \cdots & 0 \\
\vdots & \vdots  & \ddots & \vdots \\
0   & 0        & \cdots & -k_2
\end{bmatrix}
\begin{bmatrix}
\nu_{(n- \trp +1)} \\
\nu_{(n-t_r^+ +2)} \\
\vdots \\
\nu_{(n- 1)}
\end{bmatrix}
=
\begin{bmatrix}
I(t_r^+ -1) \\
I(t_r^+-2)  \\
\vdots \\
I(1)
\end{bmatrix},
\end{align}
where $\trp$ is as defined in \eqref{eq:trp}. The matrix in \eqref{eq:matrix 2 KKT} is diagonal, and has strictly negative elements along the diagonal. The right hand side of \eqref{eq:matrix 2 KKT} is similarly comprised of strictly negative elements as $I(i<\trp) <0$. Thus, the matrix equation in \eqref{eq:matrix 2 KKT} always has positive solutions for $\nu_i$, $ i \in \{n-\trp +1,...,n-1\}$. We note that the matrix in \eqref{eq:matrix 1 KKT} has an upper-triangular Toeplitz matrix structure and is invertible, so we can state $\bm{\mu} = M^{-1}\bm{b} $ exists, and therefore, the conditions for stationarity are satisfied. Moreover, we state that the non-zero dual variables stated in \eqref{eq:Nus}--\eqref{eq:Mus} were derived as the solutions to the matrix equations \eqref{eq:matrix 1 KKT}--\eqref{eq:matrix 2 KKT}.

\subsection*{Proof of (iv)}
To prove dual feasibility, we are required to show $\mu_{i} \ge 0$ for all $i \in \{1,\dots,n-\trp\}$. Given that we know both $\mu_i = 0$ and $\nu_i = -I(n - i )/k_2 \ge 0$ for $i \in \{n-\trp +1,\dots,n-1\}$, and $\nu_i = 0$ for $i \in \{1,\dots,n-\trp\}$, this leaves us to show $\mu_i \ge 0$ for $i \in \{1,\dots,n-\trp\}$. To proceed, let us define a more convenient row index $k = n - \trp - i \in \{1,\dots,n-\trp\}$ for the matrix equation in \eqref{eq:matrix 1 KKT} which starts at 1 for the bottom row and counts upwards. By back substitution, the KKT multipliers $\mu_k$ satisfy the recursion
\begin{align} \label{eq:BackSubstitution}
    k_2 \mu_{k}
    = I(\trp + k - 1) + \sum_{m=1}^{k-1} \mu_{k-m} \, I(m),
\end{align}
where $I(m) = k_1 \exp(-m/\tau_1) - k_2 \exp(-m/\tau_2)$.

\medskip

\begin{claim}\label{claim1}
Under the reparameterization \eqref{eq:Reparameterization}, the unique solution to \eqref{eq:BackSubstitution} is given by
\begin{align} \label{eq:muk polynomial}
    k_2 \mu_k = A a^{k-1} \left( \alpha Q_k(1) - \beta Q_k(z) \right),
\end{align}
where $Q_k$ is defined by the recurrence \eqref{eq:BigBoiQ}. Moreover, for all
$k \in \{1,\dots,n-\trp\}$ and all $r,z \in (0,1)$, the polynomial satisfies
\[
    Q_k(1) \ge 0.
\]
\end{claim}

\textit{Proof.} See \ref{AppB:Proof of Claim}.

\medskip

An immediate consequence of Claim~\ref{claim1} is that we can lower bound $\mu_k$ by replacing $\alpha Q_k(1)$ with the smaller, positive contribution $\beta Q_{k}(1)$,
\begin{align}
    \mu_k
    &= \frac{ A a^{k-1}}{k_2} \left( \alpha Q_k(1) - \beta Q_k(z) \right) \\
    &\ge \frac{ A a^{k-1} \beta}{k_2} \left( Q_k(1) - Q_k(z) \right)
     \;\equiv\; \frac{ A a^{k-1} \beta}{k_2} \, D_k,
\end{align}
where we define
\begin{align} \label{eq: Dk}
   D_k
   = \left( 1 - z^{k-1} \right)
   + \sum_{m=1}^{k-1} \left( r - z^m \right) D_{k-m},
   \qquad k \ge 1.
\end{align}
Proceeding via induction, we have:

\noindent \textbf{Base case.} $D_1 = 0$, $D_2 = 1-z >0$, $D_3 = (1+r)(1-z) > 0$ and $D_1<D_2<D_3$, so
\[
  D_k \ge 0 \quad \text{for } k = 1,2,3.
\]

\noindent \textbf{Induction.}
Assume that $D_m \ge 0$ for all $m < k$, where $k \ge 4$. Then the right-hand side of the recursion in \eqref{eq: Dk} for $D_k$ is monotone increasing in $r$ and monotone decreasing in $z$. Hence, $D_k$ attains its minimal value at the extremal values $r=0$ and $z=1$,
\[
  D_k \ge D_k(r=0,z =1) = 0.
\]
By induction, $D_k \ge 0$ for all $k \in \{1,\dots,n-\trp\}$.

This proves, 
\begin{align}
    \mu_k = \frac{ A a^{k-1} }{k_2}  \left( \alpha Q_k(1) - \beta Q_k(z) \right) > \frac{ A a^{k-1} \beta}{k_2} \left( Q_k(1) - Q_k(z) \right) = \frac{ A a^{k-1} \beta}{k_2} D_k \ge 0,
\end{align}
for $k = \{1,2,\dots,n-\trp\}$. Which is to say $\mu_k>0$ for $k = \{1,2,\dots,n-\trp\}$, thus, dual feasibility is satisfied, and $\bm{w^*}$ in equation \eqref{eq:TrainAsHardASPossible} is always the solution that satisfies the optimization problem in \eqref{eq:BanisterOptimizationProblem}, subject to the constraints in \eqref{con:trainingconstraintLower}--\eqref{con:trainingconstraintUpper}, for all $0<k_1<k_2$ and $0<\tau_2<\tau_1$.

\section{Proof of Claim~\ref{claim1}} \label{AppB:Proof of Claim}

\begin{proof}
We proceed in two parts. First, we verify that equation \eqref{eq:muk polynomial} is the solution to \eqref{eq:BackSubstitution}, via induction. Second, we establish the non-negativity of $Q_k(1)$ for all $k = \{1,\dots,n-\trp\}$.

\medskip
\noindent \textbf{Part 1: Proof of the closed-form solution.}
\noindent Under the reparameterization \eqref{eq:Reparameterization}, the impulse response admits the representations
\begin{align}\label{eq:ImpulseResponses}
    \frac{I(j)}{k_2} = a^j (r - z^j), \qquad
    I(t_r^+ + j) = A\, a^j (\alpha - \beta z^j).
\end{align}

\noindent \textbf{Base case.}  
For $k=1$, the back subtitution representation \eqref{eq:BackSubstitution} gives
\[
    k_2 \mu_1 = I(\trp) = A(\alpha - \beta),
\]
where we have used the reparameterized impulse response in \eqref{eq:ImpulseResponses}. On the other hand, our proposed expression in equation \eqref{eq:muk polynomial} gives
\[
    k_2 \mu_1 = A a^{0} \left( \alpha Q_1(1) - \beta Q_1(z) \right) = A(\alpha - \beta),
\]
so the formula holds for $k=1$.

\medskip
\noindent \textbf{Inductive step.}  
Assume that for some $k \ge 2$,
\begin{align} \label{eq:InductiveAssumptionCombined}
    k_2 \mu_{(k-m)} = A a^{k-m-1} \left( \alpha Q_{k-m}(1) - \beta Q_{k-m}(z) \right)
\end{align}
holds for all $m < k$. Then, using the reparameterization
\eqref{eq:Reparameterization} and the impulse response representations in \eqref{eq:ImpulseResponses},
the recursion \eqref{eq:BackSubstitution} becomes
\begin{align*}
    k_2 \mu_k
    &= I(\trp + k - 1) + \sum_{m=1}^{k-1} \mu_{k-m} I(m) \\
    &= A a^{k-1}(\alpha - \beta z^{k-1})
       + \sum_{m=1}^{k-1} a^m (r - z^m) \, k_2 \mu_{k-m}.
\end{align*}
Substituting the inductive hypothesis \eqref{eq:InductiveAssumptionCombined} into the right-hand side,
\begin{align*}
    k_2 \mu_k
    &= A a^{k-1}(\alpha - \beta z^{k-1})
       + \sum_{m=1}^{k-1} a^m (r - z^m) \, A a^{k-m-1}
         \left( \alpha Q_{k-m}(1) - \beta Q_{k-m}(z) \right) \\
    &= A a^{k-1} \Bigg[
        \alpha \left( 1 + \sum_{m=1}^{k-1} (r - z^m) Q_{k-m}(1) \right)
        - \beta \left( z^{k-1} + \sum_{m=1}^{k-1} (r - z^m) Q_{k-m}(z) \right)
      \Bigg].
\end{align*}
By the definition \eqref{eq:BigBoiQ} of $Q_k(x)$, this is exactly
\[
    k_2 \mu_k = A a^{k-1} \left( \alpha Q_k(1) - \beta Q_k(z) \right),
\]
which completes the induction and establishes \eqref{eq:muk polynomial} for all $k \ge 1$.

\medskip
\noindent \textbf{Part 2: Positivity of $Q_k(1)$.} For this part, we employ the short hand $Q_k = Q_k(1)$. The recursion \eqref{eq:BigBoiQ} becomes
\[
    Q_k = 1 + \sum_{q=1}^{k-1} (r - z^q)\, Q_{k-q}, \qquad k \ge 1.
\]

\noindent \textbf{Base cases.}  
Clearly $Q_1 = 1 \ge 0$. For $k=2$,
\[
    Q_2 = 1 + r - z \ge 1 - z \ge 0,
\]
since $z \in (0,1)$.

\medskip
\noindent \textbf{Inductive step.}  
Assume $Q_m \ge 0$ for all $m < k$, where $k \ge 3$. The right-hand side of the recursion is monotone increasing in $r$ and monotone decreasing in $z$. Therefore, $Q_k$ is minimized in the extreme case $r=0$ and $z=1$. In this worst case,
\[
    Q_1(0,1) = 1, \qquad Q_m(0,1) = 0 \quad \text{for all } m \ge 2.
\]
By the induction hypothesis, these are the smallest admissible values of $Q_m$ for $m<k$, and hence
\[
    Q_k \ge Q_k(0,1) = 0.
\]
Thus $Q_k \ge 0$ for all $k \in \{1,2,\dots,n-\trp\}$.

\medskip
Combining parts 1 and 2 completes the proof of Claim \ref{claim1}.
\end{proof}

\section{Linear Impulse Response Counterexamples} \label{sec: AppendixA2}

\begin{figure}[ht!]
     \centering
          \begin{subfigure}{0.4\textwidth}
         \centering
         \begin{overpic}[width=0.8\textwidth]{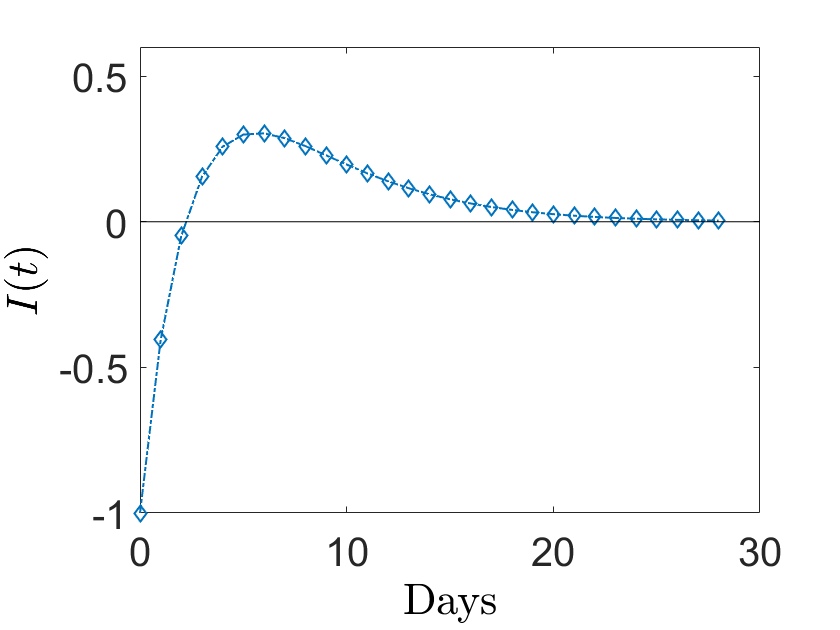}
        \end{overpic}
         \caption{ }
         \label{subfig:gAB_1}
     \end{subfigure}
     \begin{subfigure}{0.4\textwidth}
         \centering
         \begin{overpic}[width=0.8\textwidth,trim=40 0 0 0,clip]{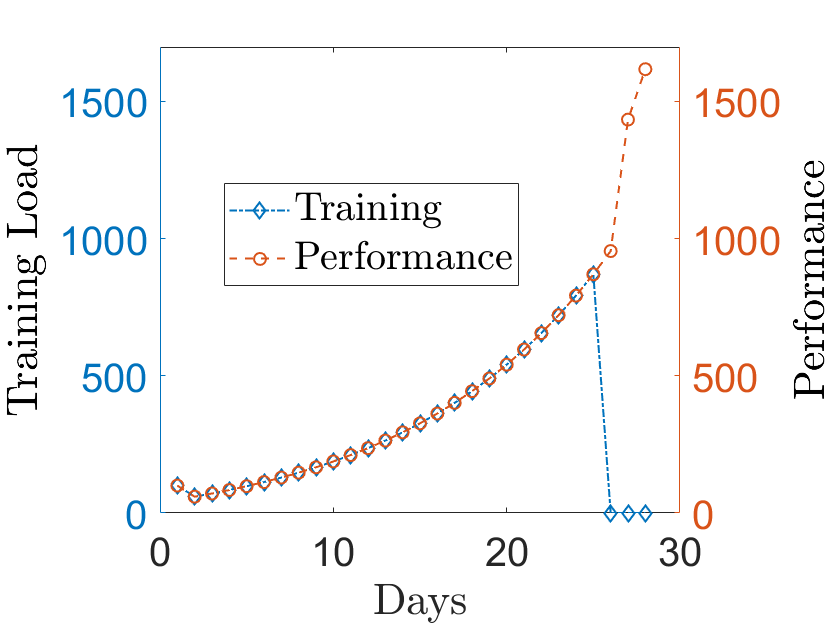}
         \put(-3,42){\rotatebox{90}{\makebox(0,0){\small Training impulse}}}
        \end{overpic}
         \caption{ }
         \label{subfig:gAB_2}
     \end{subfigure}
     \begin{subfigure}{0.4\textwidth}
         \centering
         \begin{overpic}[width=0.8\textwidth]{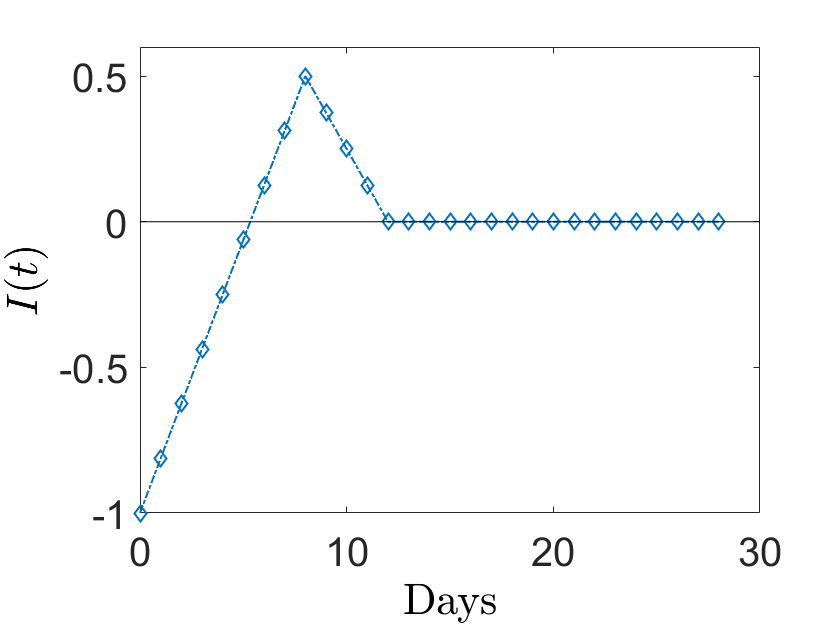}
        \end{overpic}
         \caption{ }
         \label{subfig:gAB_3}
     \end{subfigure}
     \begin{subfigure}{0.4\textwidth}
         \centering
         \begin{overpic}[width=0.8\textwidth,trim=40 0 0 0,clip]{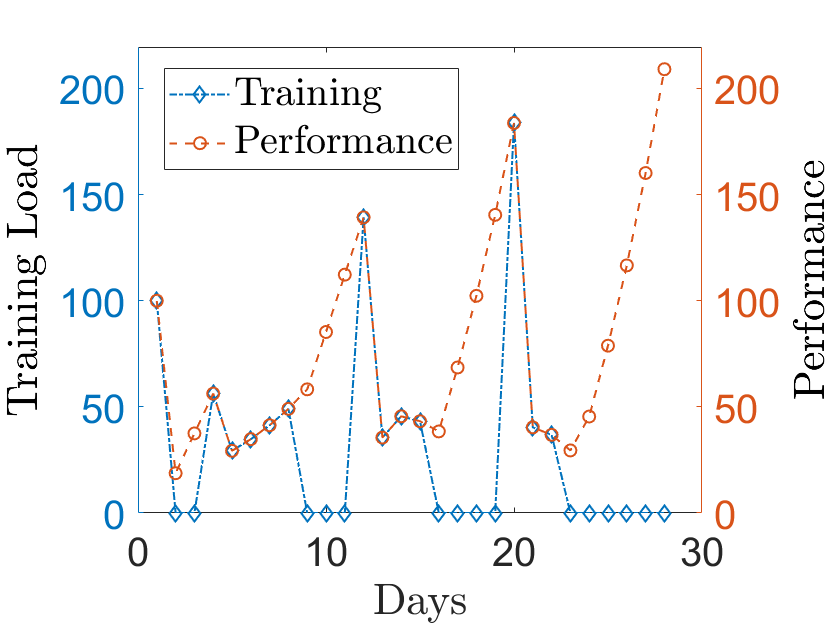}
         \put(-3,42){\rotatebox{90}{\makebox(0,0){\small Training impulse}}}
        \end{overpic}
         \caption{ }
         \label{subfig:gAB_4}
     \end{subfigure}
     \begin{subfigure}{0.4\textwidth}
         \centering
         \begin{overpic}[width=0.8\textwidth]{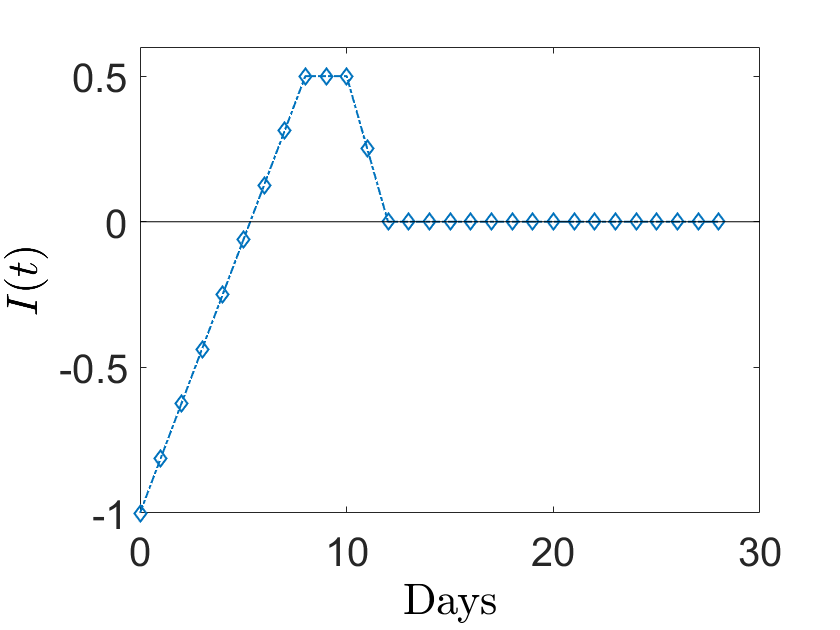}
        \end{overpic}
         \caption{ }
         \label{subfig:gAB_5}
    \end{subfigure}
    \begin{subfigure}{0.4\textwidth}
         \centering
         \begin{overpic}[width=0.8\textwidth,trim=40 0 0 0,clip]{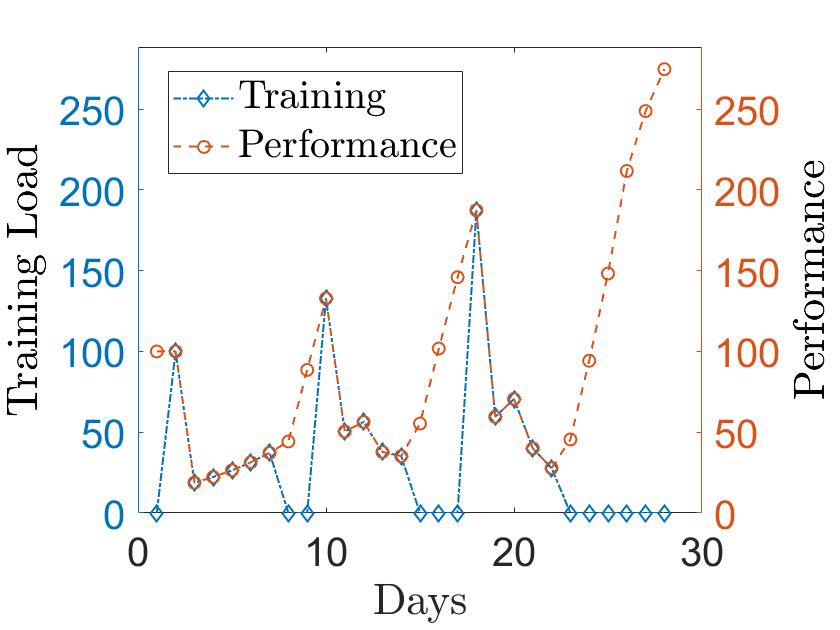}
         \put(-3,42){\rotatebox{90}{\makebox(0,0){\small Training impulse}}}
        \end{overpic}
         \caption{ }
         \label{subfig:gAB_6}
     \end{subfigure}
     \begin{subfigure}{0.4\textwidth}
         \centering
         \begin{overpic}[width=0.8\textwidth]{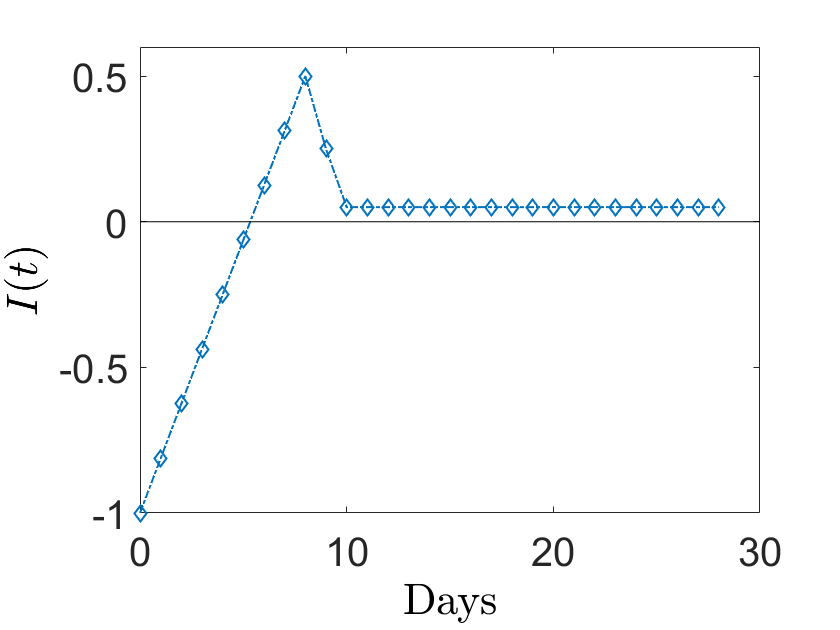}
        \end{overpic}
         \caption{ }
         \label{subfig:gAB_7}
     \end{subfigure}
     \begin{subfigure}{0.4\textwidth}
         \centering
         \begin{overpic}[width=0.8\textwidth,trim=40 0 0 0,clip]{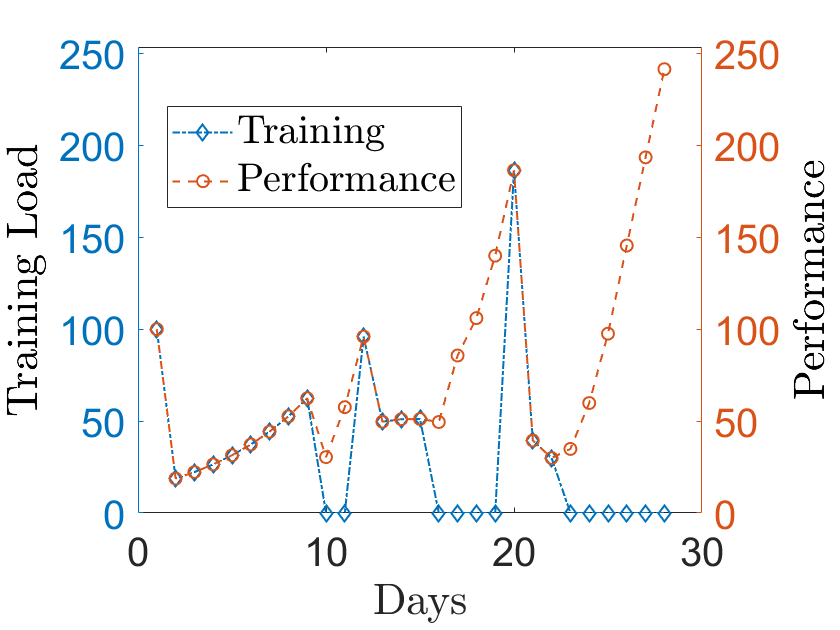}
         \put(-3,42){\rotatebox{90}{\makebox(0,0){\small Training impulse}}}
        \end{overpic}
         \caption{ }
         \label{subfig:gAB_8}
     \end{subfigure}
        \caption{(a) Banister's impulse response for the parameters $k_1  = 5, k_2      = 6, \tau_1    = 4, \tau_2    = 3$. Piecewise impulse response function as per \eqref{eq:PiecewiseIRF} for the parameter values (c) $a = 8, b = a, c = b+4, \epsilon = 0$, (e) $a = 8, b = a+2, c = b+2, \epsilon = 0$,  (g) $a = 8, b = a, c = b+2, \epsilon = 0.05$. (b)(d)(f)(h) Are the corresponding optimized training impulse profiles that maximize performance over 28 days for the linear model $P(n) = P_0 + \sum_{i=1}^{n-1} w(i)I(n-i)$, $P_0 = 100$, subject to $0\le w(i) \le P(i)$ for all $i \in \{1,\dots,n-1\}$. }
        \label{fig:Counterexamples}
\end{figure}

We present the following impulse response functions, which serve as counterexamples to the behavior of the Banister model under linear programming optimization in so much as they do not have the solution $\bm{w^*}$ in equation \eqref{eq:TrainAsHardASPossible}. Consider the following piecewise impulse response definition
\begin{align}  \label{eq:PiecewiseIRF}
    I(n) =
\begin{cases} 
-1 + \frac{3n}{2a}, & 0 \leq n < a \\
\frac{1}{2}, & a \leq n < b \\
\frac{c}{2(c-b)} -\frac{n}{2(c-b)} , & b \leq n < c \\
\epsilon , & n \geq c,
\end{cases}
\end{align}
where $a,b,c,\epsilon$ are parameters of the impulse response function. A graphical representation of this for three different sets of parameters can be observed in Fig.~\ref{subfig:gAB_3}-\ref{subfig:gAB_5}-\ref{subfig:gAB_7}. Employing these in the place of Banister's impulse response in equation \eqref{Eq:OGBanisterDiscrete}, subject to $0\le w(i) \le P(i)$ for all $i \in \{1,\dots,n-1\}$, the linear programming optimization results can be observed in Fig.~\ref{subfig:gAB_4}-\ref{subfig:gAB_6}-\ref{subfig:gAB_8}. In these illustrative training impulse profiles, we no longer see the ``train every day as hard as we can up to $(n-\trp)$'', such as is the case for the Banister model, Fig.~\ref{subfig:gAB_1}-\ref{subfig:gAB_2}. Instead, we see that the linear programming optimization suggests periodization, which could be qualitatively described as: train for several days to build up our fitness at the cost of fatigue, rest to realize the positive change in performance, use that increased performance to train more than before, and so on in cycles of a similar fashion.

\label{Bibliography}
\bibliographystyle{plainnat}
\bibliography{PhD_Start_Bib.bib}

\end{document}